\documentclass{aa}

\usepackage{graphicx}
\usepackage{hyperref}
\usepackage{booktabs}
\usepackage{indentfirst}
\usepackage{placeins}
\usepackage{makecell}
\usepackage{siunitx}
\usepackage{cuted}
\usepackage[singlelinecheck=false 
,labelsep=period,labelfont=bf]{caption}
\usepackage{txfonts}
\usepackage{footnote}
\usepackage{supertabular}
\begin{document}

   \title{Distance estimates for AGB stars from parallax measurements\thanks{Table C.1 is only available in electronic form
at the CDS via anonymous ftp to cdsarc.u-strasbg.fr (130.79.128.5)
or via \url{http://cdsweb.u-strasbg.fr/cgi-bin/qcat?J/A+A/}}}

   \author{M. Andriantsaralaza
          \inst{1}
          \and
          S. Ramstedt
          \inst{1}
          \and
          W. H. T. Vlemmings \inst{2}
          \and
          E. De Beck \inst{2}
          }

         \institute{Theoretical Astrophysics, Division for Astronomy and Space Physics, Department of Physics and Astronomy, Uppsala University, Box 516, SE-751 20 Uppsala, Sweden\\
              \email{miora.andriantsaralaza@physics.uu.se}
        \and
        Department of Space, Earth and Environment, Chalmers University of Technology, Onsala Space Observatory, 439 92 Onsala, Sweden
             }

   \date{Received 29 March 2022 / Accepted 24 August 2022 }

 
  \abstract
   {Estimating the distances to asymptotic giant branch (AGB) stars using optical measurements of their parallaxes is not straightforward because of the large uncertainties introduced by their dusty envelopes, their large angular sizes, and their surface brightness variability.}
   {This paper aims to assess the reliability of the distances derived with \textit{Gaia} DR3 parallaxes for AGB stars, and provide a new distance catalogue for a sample of $\sim200$ nearby AGB stars.}
   {We compared the parallaxes from \textit{Gaia} DR3 with parallaxes measured with maser observations with very long baseline interferometry (VLBI) to determine a statistical correction factor for the DR3 parallaxes using a sub-sample of $33$ maser-emitting oxygen-rich nearby AGB stars. We then calculated the distances of a total of $\sim200$ AGB stars in the DEATHSTAR project using a Bayesian statistical approach on the corrected DR3 parallaxes and a prior based on the previously determined Galactic distribution of AGB stars. We performed radiative transfer modelling of the stellar and dust emission to determine the luminosity of the sources in the VLBI sub-sample based on the distances derived from maser parallaxes, and derived a new bolometric period-luminosity relation for Galactic oxygen-rich Mira variables.}
   {We find that the errors on the \textit{Gaia} DR3 parallaxes given in the \textit{Gaia} DR3 catalogue are underestimated by a factor of $5.44$ for the brightest sources ($G<8$~mag). Fainter sources ($8\leq G<12$) require a lower parallax error inflation factor of $2.74$. We obtain a \textit{Gaia} DR3 parallax zero-point offset of $-0.077\,$mas for bright AGB stars. The offset becomes more negative for fainter AGB stars. After correcting the DR3 parallaxes, we find that the derived distances are associated with significant, asymmetrical errors for more than 40\,\% of the sources in our sample. We obtain a PL relation of the form $M_\mathrm{bol} = (-3.31\pm{0.24})\,[\mathrm{log}\, P - 2.5] + (-4.317\pm0.060)$ for the oxygen-rich Mira variables in the Milky Way. A new distance catalogue based on these results is provided for the sources in the DEATHSTAR sample. }
   {The corrected \textit{Gaia} DR3 parallaxes can be used to estimate distances for AGB stars using the AGB prior, but we confirm that one needs to be careful when the uncertainties on  parallax measurements are larger than 20\,\%, which can result in model-dependent distances and source-dependent offsets. We find that a RUWE (re-normalised unit weight error) below 1.4 does not guarantee reliable distance estimates and we advise against the use of only the RUWE to measure the quality of \textit{Gaia} DR3 astrometric data for individual AGB stars.}


   \maketitle
%

\section{Introduction}

Distance is one of the most fundamental parameters in astronomy which lies at the basis of the analysis and interpretation of astronomical data. The \textit{Gaia} mission \citep{Gaia2016science} aims to provide accurate measurements of the position, the parallax, and the proper motions of about 1\,\% of the stars in the Milky Way with accuracies 100 times better than its predecessor \textit{Hipparcos}. Its astrometric instrument covers wavelengths between 330 and 1050 nm, defining the photometric $G$ band \citep{Carrasco2016, Gaia2016science,DR12016,vanlee2017}. The third \textit{Gaia} data release (DR3), recently published by \citet[][]{GaiaDR32022}, corresponds to an observing time of 34 months. The corresponding astrometry was published in an early third data release \citep[][hereafter eDR3]{eDR32021}, with nominal parallax uncertainties of $0.02-0.03$ mas for $G<15$, $0.07$ mas at $G=17$, $0.5$ mas at $G=20$, and $1.3$ mas at $G=21$ mag. We note that the values of all the parameters relevant to this work (e.g. parallax, parallax error,  $G$ magnitude) are identical in the \textit{Gaia} eDR3 and DR3 catalogues for the sources discussed in this paper.
  
Determining the distances to asymptotic giant branch (AGB) stars using parallaxes measured with optical telescopes such as \textit{Gaia} is, however, not a simple task. Comparative studies such as the analysis by \citet{Xu2019} based on \textit{Gaia} data release 2 \citep[DR2;][]{DR22018} show that the precision of the \textit{Gaia} parallaxes depends on the colour of the star: the redder the star, the larger the errors. This is because red stars are usually larger, show more surface brightness variation, and have more dust. This is true for AGB stars. The AGB phase is the evolutionary stage at which low-to-intermediate-mass stars lose mass through slow and massive stellar winds, with mass-loss rates reaching up to $10^{-4}\,\mathrm{M}_\odot\, \mathrm{yr}^{-1}$ \citep{review2018}. The material ejected from the star forms a large envelope mainly consisting of molecules and dust, called the circumstellar envelope (CSE). Therefore, interstellar and circumstellar dust both contribute to making AGB stars nearly invisible in the optical range of the electromagnetic spectrum. In addition, AGB stars are large objects, having angular sizes on the order of, or larger than, their respective parallaxes for the nearby ones \citep[e.g.][]{Chiavassa2020}. Moreover, observations and simulations of AGB stars show that they possess large convective cells on their surfaces, which can shift the photocentre and thus introduce additional uncertainties to the measured parallaxes \citep{Chiavassa2018}. Furthermore, observations of bright sources can lead to instrumental saturation, resulting in less accurate astrometric measurements \citep{El-Badry2021}. This applies to AGB stars, as they are intrinsically bright objects.
 
An alternative method for parallax measurement consists of observing maser emission using very long baseline interferometry (VLBI). This method can yield a parallax precision of $\sim 10 \,\mu$as, comparable to or even better than \textit{Gaia} \citep[e.g.][]{Reid2014}. However, it can only be used to determine the distance of known maser-emitting sources, which represent only a small number of AGB stars.  Comparing  the VLBI parallaxes with the \textit{Gaia} DR3 parallaxes is therefore critical to get a better idea of the actual errors on the latter \citep{VanLan2018,Xu2019} in order to infer better distance estimates for a large sample of AGB stars.
 
The main objective of this work is to determine the distances of $\sim$\,200 nearby AGB stars that are part of the DEATHSTAR sample, described in Sect.~\ref{sec: sample}. To attain this goal, this paper is divided into four main parts. First, the \textit{Gaia} DR3 parallaxes were calibrated using a sub-sample of maser-emitting oxygen-rich AGB stars that have independent parallax measurements obtained using VLBI techniques (Sect.~\ref{sec: correction}). We then calculated the distances of the sources in the DEATHSTAR sample using the newly corrected \textit{Gaia} DR3 parallaxes following a Bayesian approach (Sect.~\ref{sec: Bayesian distance}). In Sect.~\ref{sec: PL} we present an alternative method to determine the distances of Mira variables in the DEATHSTAR sample: a new bolometric period-luminosity (PL) relation based on the aforementioned independent maser parallax measurements of the sub-sample of VLBI sources. Finally, we compiled a new distance catalogue for the $\sim\,$200 nearby AGB stars in the DEATHSTAR sample in Sect.~\ref{sec: cat}, based on the results in Sects.~\ref{sec: Bayesian distance} and \ref{sec: PL}, and using alternative distance determination methods in the literature, when needed. Section~\ref{sec: conclusion} closes the paper with a summary and conclusions.
\section{The sample}
\label{sec: sample}
\begin{figure}[t]
\includegraphics[scale=0.315]{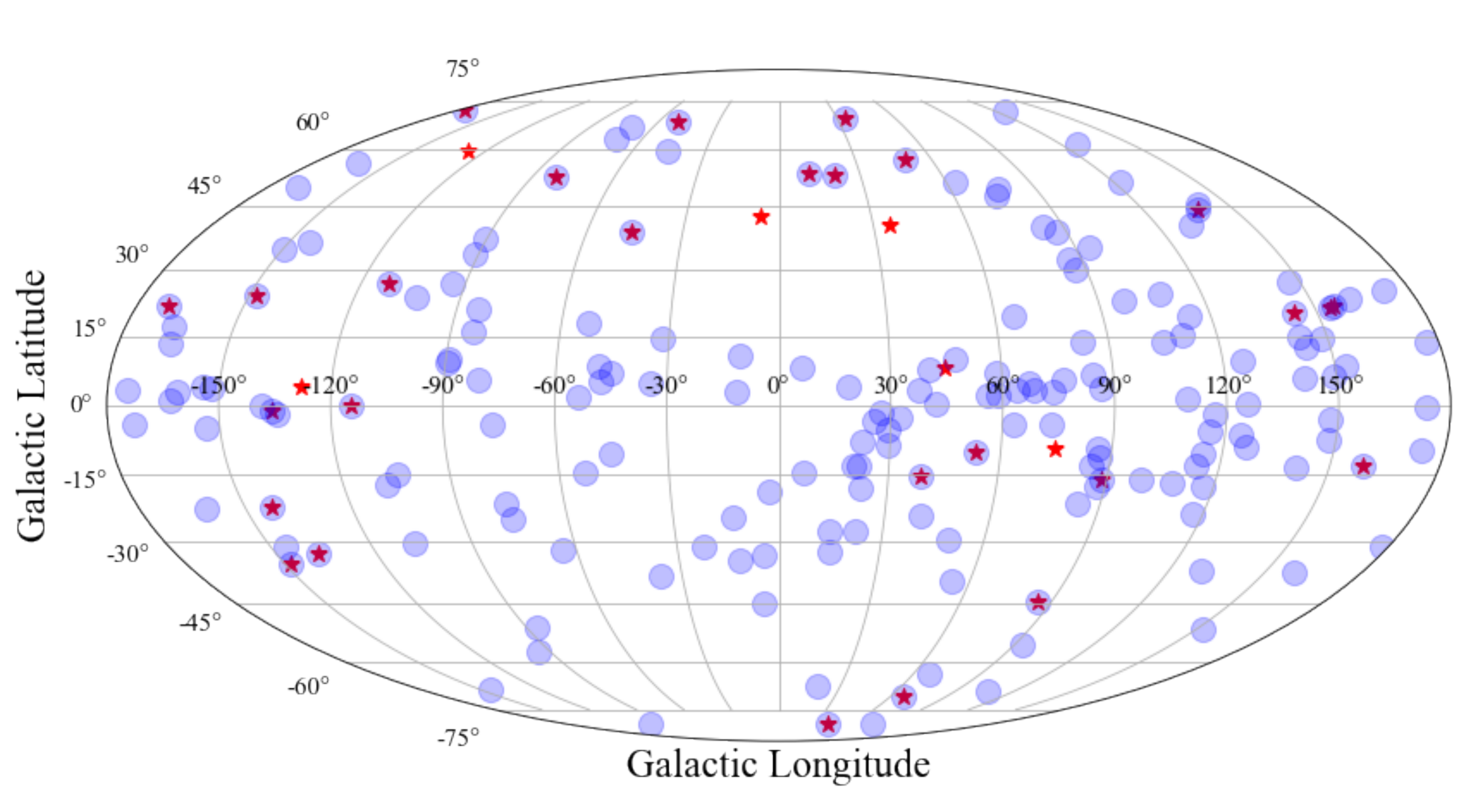}
\caption{Position of the stars in our sample. The circles represent the stars part of the DEATHSTAR sample while the red stars show the AGB stars in the VLBI sample. }
\label{fig: galactic distribution}
\end{figure}
The DEATHSTAR\footnote{\url{www.astro.uu.se/deathstar}} sample consists of $\sim$~200 nearby AGB stars. The first publications of the DEATHSTAR project present the CO observations of $\sim70$ southern sources \citep{Death1,Death2}. The source selection and completeness of the full DEATHSTAR sample is discussed in \citet{Death1}. The C-type stars (C/O\,$>$\,1) are all brighter than 2\,mag in the $K$ band and were taken from \citet{schoier2001}. The M-type stars (C/O\,$<$\,1) were collected from either the General Catalogue of Variable Stars \citep[GCVS;][non-Miras]{GCVS} or \citet[][Miras]{Gonzalez2003}. The selection criteria for the M-type stars taken from the GCVS are a quality flag 3 (high quality) in the \textit{IRAS} 12, 25, and 60\,$\mu$m bands, and a 60\,$\mu$m flux $\geq3$\,Jy. The S-type stars (C/O\,$\simeq$\,1) were also selected based on the quality of their \textit{IRAS} flux measurements at 12, 25, and 60\,$\mu$m, and on the presence of Tc in their spectra. They were collected from \citet{catalogueS} and \citet{jorissen1998}.
About $\sim93$\,\% of the stars in the DEATHSTAR sample have their parallaxes measured by \textit{Gaia} and are in eDR3 and DR3.
{
\renewcommand{\arraystretch}{1.05}
\begin{table*}[t]
 
\caption{Properties of the VLBI sources. }
  \begin{center}
    \begin{tabular}{lccccccccccc}
    \toprule
     Source & Var&$P$& $\varpi_\mathrm{VLBI}$ & $\sigma_{\varpi} ^\mathrm{VLBI}$ & $\varpi_\mathrm{DR3}$ & $\sigma_\varpi ^\mathrm{DR3}$ & excess noise & RUWE & $G$ & $G_{BP}$& $G_{RP}$ \\
      & &[days]& [mas]&[mas] & [mas] & [mas] &[mas] & &[mag]&[mag]&[mag]\\
     \midrule
AP Lyn & M & 730 & 2.00 &0.04& 2.02&0.12& 1.07&1.60
&8.68&13.88&6.94\\
BX Cam &M& 486 & 1.73 & 0.03 & 1.76 & 0.1 & 1.06 & 1.4 & 10.06 & 16.04 & 8.37 \\
BX Eri &M& 165 & 2.12 & 0.1 & 2.35 & 0.06 & 0.53 & 1.0 & 6.77 & 10.1 & 5.15 \\
FV Boo &SR& 313 & 0.97 & 0.06 & 1.01 & 0.09 & 0.79 & 1.4 & 10.59 & 15.05 & 8.86 \\
HS UMa &LB& - & 2.82 & 0.1 & 3.2 & 0.1 & 0.56 & 1.4 & 6.08 & 9.34 & 4.49 \\
HU Pup &SRa& 238 & 0.31 & 0.04 & 0.29 & 0.03 & 0.28 & 1.3 & 7.02 & 8.88 & 5.74 \\
NSV 17351 &M& 680 & 0.25 & 0.01 & 0.09 & 0.15 & 1.31 & 2.1 & 12.57 & 18.2 & 11.15 \\
OZ Gem &M& 598 & 0.81 & 0.04 & 0.46 & 0.33 & 2.41 & 3.3 & 13.84 & 18.48 & 11.95 \\
QX Pup &M& 551 & 0.61 & 0.03 & 0.03 & 0.16 & 0.0 & 1.0 & 18.27 & 18.68 & 16.34 \\
R Aqr & M&387 & 4.59 & 0.24 & 2.59 & 0.33 & 1.33 & 2.0 & 6.71 & 10.76 & 4.87 \\
R Cnc &M& 357 & 3.84 & 0.29 & 3.94 & 0.18 & 1.13 & 2.0 & 6.53 & 10.68 & 4.92 \\
R Hya & M&380 & 7.93 & 0.18 & 6.74 & 0.46 & 2.61 & 2.9 & 3.15 & 6.84 & 2.33 \\
R Peg & M&378.1 & 2.76 & 0.28 & 2.63 & 0.12 & 0.63 & 1.3 & 7.6 & 12.06 & 5.84 \\
R UMa & M&301.62 & 1.97 & 0.05 & 1.75 & 0.09 & 0.71 & 1.0 & 7.92 & 11.61 & 6.41 \\
RR Aql & M&396.1 & 2.44 & 0.07 & 1.95 & 0.11 & 0.84 & 1.5 & 8.43 & 13.15 & 6.81 \\
RT Vir & SRb&157.9 & 4.42 & 0.13 & 4.14 & 0.23 & 1.33 & 1.3 & 5.09 & 8.42 & 3.52 \\
RW Lep &SRa& 149.9 & 1.62 & 0.16 & 2.54 & 0.08 & 0.64 & 1.2 & 7.17 & 10.74 & 5.56 \\
RX Boo &SRb& 158 & 7.31 & 0.5 & 6.42 & 0.23 & 1.59 & 1.7 & 4.37 & 7.84 & 2.94 \\
S CrB &M& 360.26 & 2.39 & 0.17 & 2.6 & 0.11 & 1.09 & 1.6 & 6.86 & 11.19 & 5.09 \\
S Crt & SRb&155 & 2.33 & 0.13 & 2.06 & 0.1 & 0.63 & 1.4 & 6.34 & 9.26 & 4.78 \\
S Ser & M&371.84 & 1.25 & 0.04 & 0.77 & 0.13 & 1.12 & 3.5 & 8.41 & 12.36 & 6.74 \\
SV Peg & SRb&144.6 & 3.0 & 0.06 & 2.59 & 0.17 & 1.2 & 4.4 & 5.67 & 9.04 & 4.07 \\
SY Aql &M& 355.92 & 1.1 & 0.07 & 1.07 & 0.09 & 0.81 & 1.4 & 9.36 & 13.67 & 7.83 \\
SY Scl &M& 411 & 0.75 & 0.03 & 0.52 & 0.12 & 0.93 & 2.0 & 9.74 & 13.98 & 7.96 \\
T Lep &M& 372 & 3.06 & 0.04 & 3.09 & 0.1 & 0.95 & 2.3 & 6.91 & 11.32 & 5.24 \\
U Her &M& 404 & 3.76 & 0.27 & 2.36 & 0.08 & 0.67 & 1.3 & 6.91 & 11.12 & 5.21 \\
U Lyn &M& 433.6 & 1.27 & 0.06 & 1.01 & 0.08 & 0.61 & 1.3 & 8.49 & 12.84 & 6.75 \\
UX Cyg &M& 569 & 0.54 & 0.06 & 0.7 & 0.09 & 1.0 & 3.0 & 10.0 & 14.39 & 8.47 \\
V637 Per &SR& - & 0.94 & 0.02 & 0.85 & 0.1 & 0.73 & 1.3 & 9.04 & 12.43 & 7.46 \\
V837 Her & M & 514 & 1.09 & 0.01 & 0.18 & 0.10 & 0.85 & 1.2 & 10.74 & 16.55& 9.05\\
W Leo &M& 391.75 & 1.03 & 0.02 & 0.88 & 0.11 & 0.65 & 1.0 & 9.83 & 14.48 & 8.21 \\
X Hya &M& 299.5 & 2.07 & 0.05 & 2.53 & 0.11 & 0.68 & 1.6 & 7.88 & 11.66 & 5.88 \\
Y Lib &M& 277 & 0.86 & 0.05 & 0.83 & 0.08 & 0.6 & 1.6 & 9.76 & 13.68 & 8.09 \\
    \hline
    \multicolumn{9}{l}{\footnotesize Var: variability type.} \\
    \multicolumn{9}{l}{\footnotesize M: Mira, SRa/b: semi-regular a or b.} \\
    \multicolumn{9}{l}{\footnotesize LP: long period variable, U: unknown.} \\
    \multicolumn{9}{l}{\footnotesize $P$: period from VSX\footnotemark \citep{Watson2021}.}\\
    \end{tabular}
    \label{table: dist VLBI}
  \end{center}
\end{table*}
}
\footnotetext{Variable star index; \url{https://www.aavso.org/vsx/index.php}}
\section{Correcting the \textit{Gaia} parallaxes}
\label{sec: correction}
\subsection{VLBI parallax measurement}
Long-period variables, including AGB stars, can have strong and compact maser emission (e.g. OH, SiO, H$_2$O masers) that can be tracked to measure parallaxes \citep[e.g.][]{Vlemmings2002,Vlemmings2003,Vlemmings2007, Nyu2011,Kamezaki2016, Nakagawa2014PL, Chibueze2020, VERACol}. Parallax measurements with VLBI make use of the phase-referencing method where the maser-emitting source is monitored once per month, for example, simultaneously with a bright reference source over the course of one or more years. The parallax is obtained by fitting the offsets in the position of the maser spot as a function of time. The errors on the measured parallaxes include both systematic, such as atmospheric calibration errors, and thermal errors.  The main assumption needed to measure the parallaxes with this technique is on the motion of the maser features with respect to the star. For some sources, a maser that occurs in the outer CSE on the direct line of sight to the star amplifies the radio emission from the stellar radiophotosphere, producing the so-called amplified stellar image. In this case, the motion of the maser directly reflects the motion of the AGB star \citep{Vlemmings2003}.  Alternatively, one can assume that the maser features follow linear motions in the shell, as in \citet{Vlemmings2007}. In most cases, the motions of maser spots are small \citep{Vlemmings2003}, thus parallaxes measured with VLBI astrometry are highly accurate. For AGB stars in particular, parallaxes obtained from maser observations are more robust than \textit{Gaia} DR3 parallaxes, which are measured in the optical, as they are not affected by dust obscuration. Furthermore, tracking the maser spots overcomes problems related to stellar variability \citep[e.g.][]{Vlemmings2003,Vlemmings2007}. Therefore, the more precise VLBI parallaxes can be used to calibrate the \textit{Gaia} DR3 parallaxes and their uncertainties.
 \subsection{VLBI sample}
 \label{sec: VLBI sample}
To calibrate the \textit{Gaia} DR3 parallaxes, we used existing VLBI parallax measurements of maser-emitting oxygen-rich AGB stars in the literature. We first considered the AGB stars in the sample of the VLBI Exploration of Radio Astrometry or  VERA\footnote{\url{https://www.miz.nao.ac.jp/veraserver/related/index-e.html}} catalogue \citep{VERACol}.
The VERA survey consists of four 20-m radio telescopes targeting $\mathrm{H}_2$O and SiO maser emission at $22$ and $43$ GHz, yielding maps with an angular resolution reaching up to 1.2  and 0.7 mas, respectively. The VERA catalogue comprises 99 objects out of which 29 are labelled as AGB stars. Out of these, 3 are proto-planetary nebulae/post-AGB stars and 26 are AGB stars, according to the SIMBAD\footnote{\url{http://simbad.u-strasbg.fr/simbad/}} database. Three additional VERA sources were published by \citet{Chibueze2020} and \citet{Yan2022}. We also considered the maser-emitting AGB stars in the sample presented in \citet{Xu2019} which was compiled from the results of \cite{Vlemmings2003,Kurayama2005,Vlemmings2007}, and \citet{Zhang2017}. Out of the 13 sources labelled as AGB stars that do not overlap with the VERA sample in the \citet{Xu2019} sample, 6 are actually known supergiants or hypergiants. Moreover, as our aim was to use the VLBI parallaxes as calibrators for the \textit{Gaia} DR3 parallaxes, we disregarded the sources whose VLBI parallax measurements have a signal-to-noise ratio lower than 5. This is the case for BW~Cam, R~Cas, and W~Hya. Therefore, our VLBI sample consists of an aggregate of 33 maser-emitting AGB stars (28 VERA, 5 from \citealt{Xu2019}).

The Galactic distribution and the overlap of the DEATHSTAR and the VLBI samples are shown in Fig.~\ref{fig: galactic distribution}. The properties of the VLBI sources are listed in Table~\ref{table: dist VLBI}. We retrieved the \textit{Gaia} DR3 parallaxes and parallax uncertainties, $\varpi_\mathrm{DR3}$ and $\sigma_\varpi ^\mathrm{DR3}$, respectively, for the sources in the VLBI sample, along with their $G$, $G_{BP}$, and $G_{RP}$ magnitudes from the \textit{Gaia} online archive\footnote{\url{https://gea.esac.esa.int/archive/}}. The astrometric excess noise and the RUWE (re-normalised unit weight error), also taken from the \textit{Gaia} archive, measure discrepancies in photocentric motions, and therefore, quantify the goodness-of-fit of the \textit{Gaia} DR3 astrometric data. The period $P$ and variability of the stars were taken from the Variable Star Index or VSX online tool \citep{Watson2021}.

\subsection{\textit{Gaia} DR3 parallax}
\subsubsection{G magnitude and colour}
\label{sec:Gmag color}
\begin{figure*}[ht]
\centering
\includegraphics[scale=0.21]{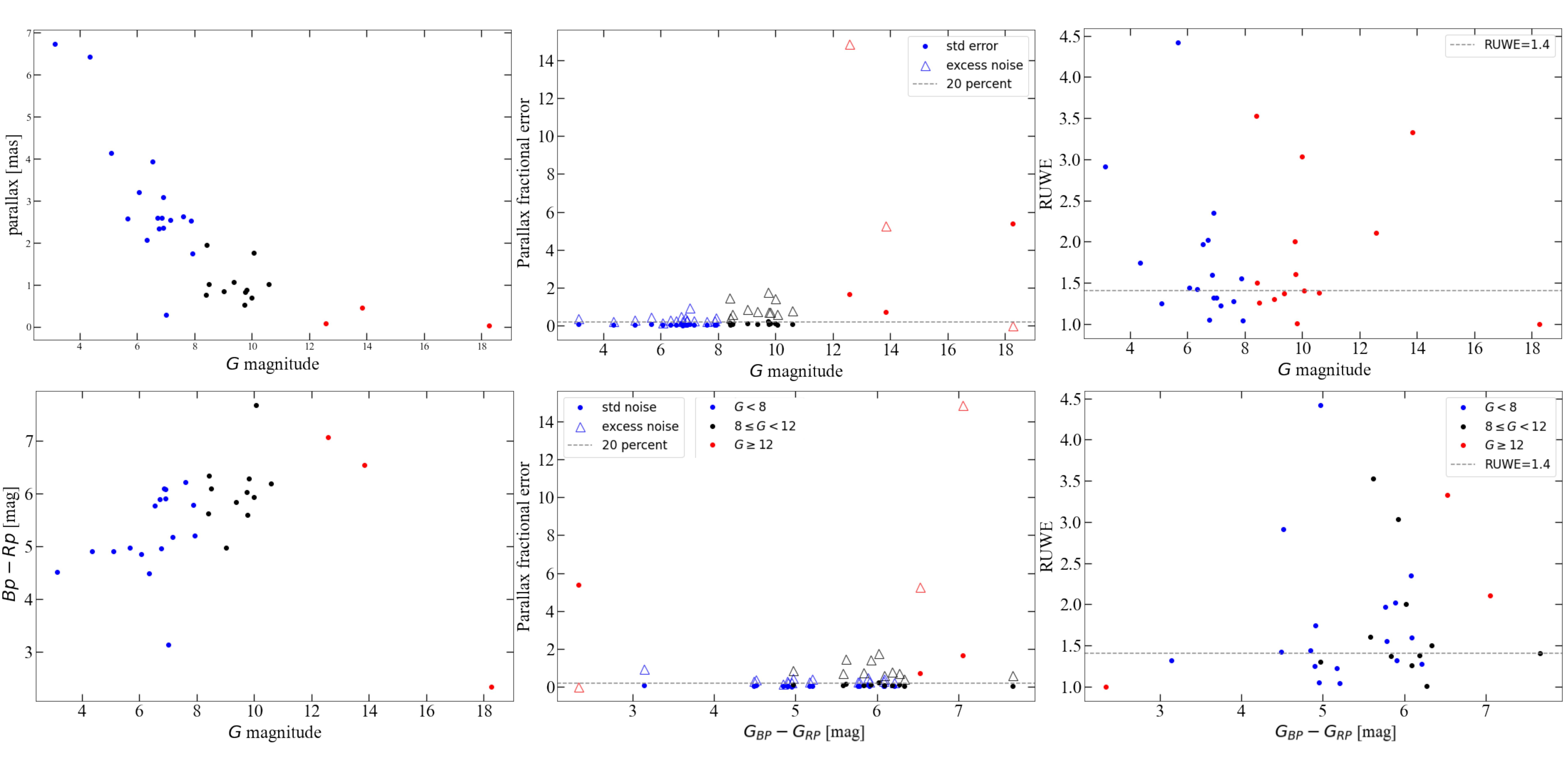}
 
\caption{Dependence of the standard noise, the astrometric excess noise, and the RUWE of the \textit{Gaia} DR3 parallax on the $G$ magnitude and the colour $G_{BP} - G_{RP}$ for the sources in the VLBI sample.}
\label{fig: VLBI Gaia parallax}
\end{figure*}

\begin{figure*}[ht]
\centering
\includegraphics[scale=0.22]{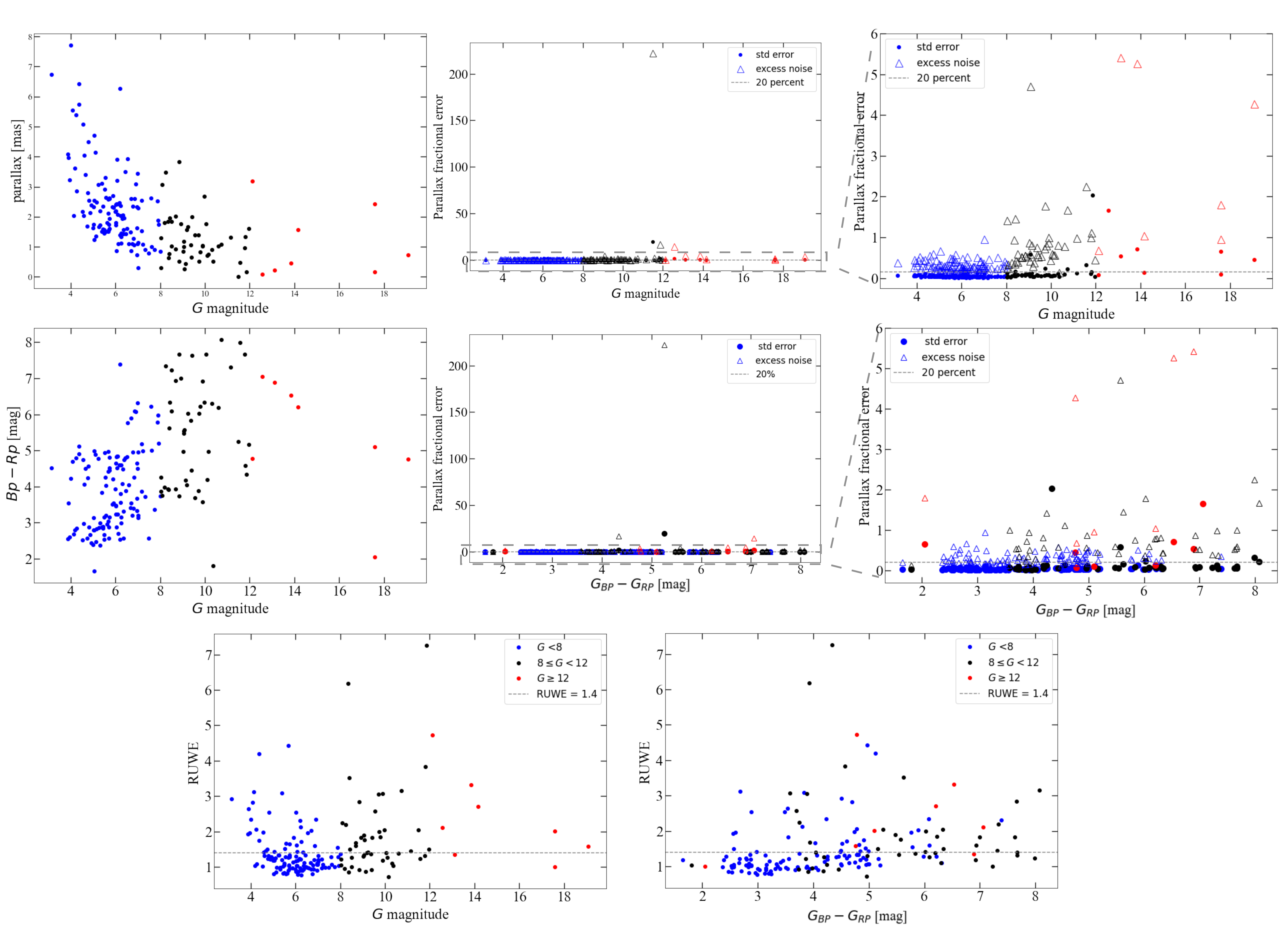}
 
\caption{Dependence of the standard noise, the astrometric excess noise, and the RUWE of the \textit{Gaia} DR3 parallax on the $G$ magnitude and the colour $G_{BP} - G_{RP}$ for the sources in the DEATHSTAR sample.}
\label{fig: DEATHSTAR Gaia parallax}
\end{figure*}

We first investigated the dependence of the \textit{Gaia} DR3 parallax uncertainties and the parameters that represent the goodness-of-fit of the astrometric data on the $G$ magnitude, and the colour $G_{BP} - G_{RP}$ for the sources in the VLBI sample (Fig.~\ref{fig: VLBI Gaia parallax}) and in the DEATHSTAR sample (Fig.~\ref{fig: DEATHSTAR Gaia parallax}). It is apparent in Figs.~\ref{fig: VLBI Gaia parallax} and ~\ref{fig: DEATHSTAR Gaia parallax} that the fainter the star, the redder it is and the smaller its parallax, as expected. However, this behaviour breaks down for the faintest stars ($G\geq12$~mag). The nominal fractional parallax error seems mostly reasonable (below 20\,\%) for sources brighter than $G=8$~mag, but shows a slight increase between $8$ and $12\,G$~magnitudes. For the three faintest sources in the VLBI sample, the standard \textit{Gaia} DR3 parallax error is larger than 75\,\%  (Fig.~\ref{fig: VLBI Gaia parallax}), while a wider scatter is observed for the faintest stars in the DEATHSTAR sample (Fig.~\ref{fig: DEATHSTAR Gaia parallax}). The astrometric excess noise worsens with $G$~magnitude, with an obvious general increase above $G=8$~mag. It does not show a strong dependence on the colour. The RUWE does not strongly correlate with neither the colour nor the $G$ magnitude.
\\We divided the sources in the VLBI and the DEATHSTAR samples into three categories according to their $G$  magnitudes: (1) $G<8$, (2) $8 \leq G<12$, and (3) $G\geq12$ mag. This division is mainly based on the behaviour of their parallax fractional error and the goodness of the \textit{Gaia} DR3 astrometric measurements measured by their astrometric excess noise, as seen in Figs.~\ref{fig: VLBI Gaia parallax} and \ref{fig: DEATHSTAR Gaia parallax}.
\subsubsection{Zero-point offset and error inflation factor}
\begin{figure}[t]
\centering
\includegraphics[scale=0.3]{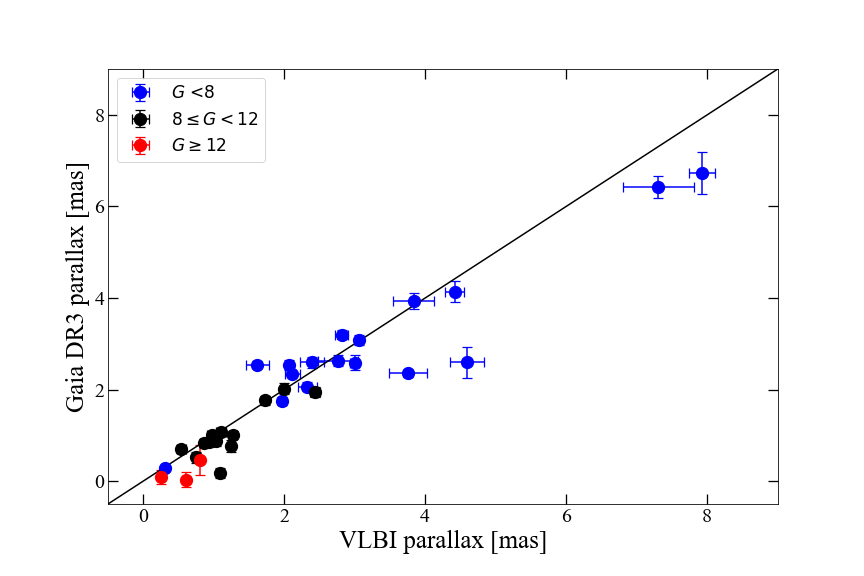}
 
\caption{Comparison between the VLBI and \textit{Gaia} DR3 parallaxes. The solid line represents the 1-to-1 relation .}
\label{fig: VLBI vs Gaia}
\end{figure}

Figure~\ref{fig: VLBI vs Gaia} shows a direct comparison of the VLBI and the \textit{Gaia} DR3 parallaxes, $\varpi_\mathrm{VLBI}$ and $\varpi_\mathrm{DR3}$, respectively, for the sources in the VLBI sample. A rather good agreement is observed for the stars with parallaxes lower than 4 mas, with increasing scatter and uncertainties at higher parallax values, corresponding to the brightest sources, along both axes. Figure~\ref{fig: VLBI vs Gaia} also shows that the nominal noise of the \textit{Gaia} DR3 parallaxes, $\sigma_\varpi ^\mathrm{DR3}$,  is smaller than the VLBI parallax uncertainties $\sigma_\varpi ^\mathrm{VLBI}$. However, the uncertainties of the parallaxes measured with \textit{Gaia} are known to be underestimated, especially for variable objects such as AGB stars. The previously mentioned astrometric excess noise was introduced in DR2 as the extra uncertainty that must be added in quadrature to the nominal noise to obtain a statistically acceptable astrometric solution for \textit{Gaia} DR2 measurements \citep{Lindegren2012}. A comparative study was conducted by \citet{VanLan2018} using \textit{Gaia} DR2 and VLBI measurements for a number of sources including Mira variables, pre-main sequence stars, and binary pulsars. \citet{VanLan2018} attribute the large residual in the \textit{Gaia} parallaxes to the effects of the properties, such as the colour and surface brightness distribution, of the AGB stars in their sample. Their study confirms that adding the astrometric excess noise of the \textit{Gaia} measurements to the nominal noise equalises the \textit{Gaia} DR2 parallaxes with the more robust VLBI measurements.

\begin{figure*}[t]
\centering
\includegraphics[scale=0.315]{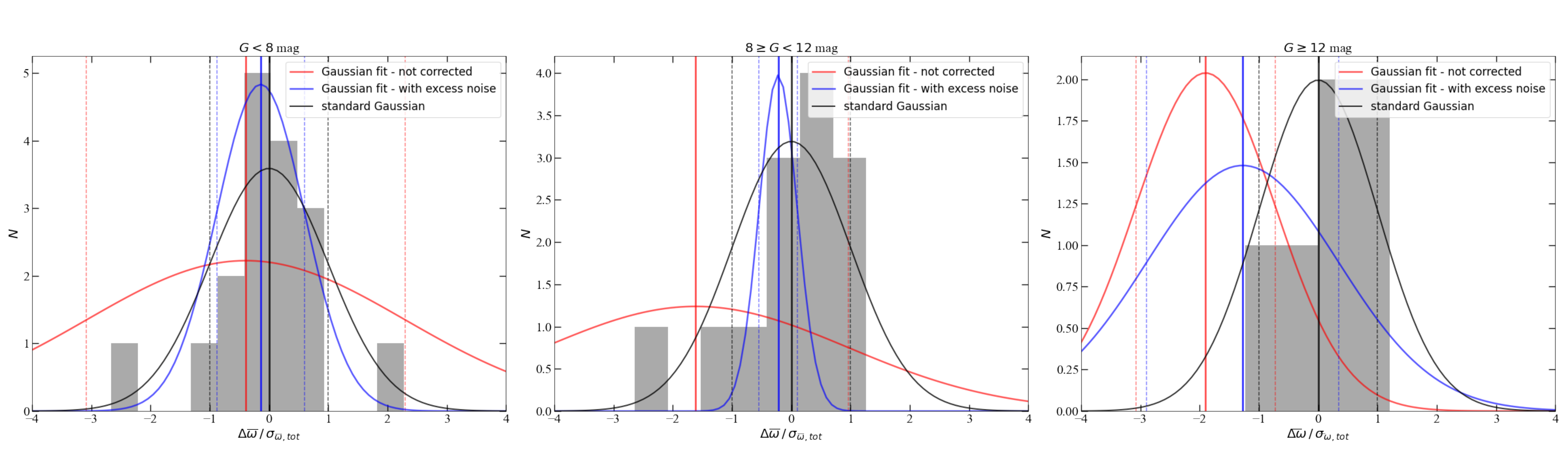}
 
\caption{Gaussian fittings of the parallax difference distribution  $\Delta \omega \, / \, \sigma_{\varpi, tot}$
for the three $G$~magnitude categories. The distributions were normalised by the quadratically
summed errors of the VLBI error with either: the DR3 parallax standard error without any correction (${\sigma_{\varpi, tot} = \sqrt{ (\sigma_\varpi ^\mathrm{DR3} ) ^2+(\sigma_\varpi ^{VLBI} )^2}}$, red line), or with the astrometric excess noise (${\sigma_{\varpi, \mathrm{tot}}= \sqrt{(\sigma_\varpi ^\mathrm{DR3}) ^2+\sigma^2 _\mathrm{\small{excess\,noise}} \, + (\sigma_\varpi ^\mathrm{VLBI}) \,^2}}$, blue
line), or with the DR3 parallax standard error
inflated by the EIF listed in Table~\ref{table: EIF and ZPO} and the parallax corrected for the ZPO (${\sigma_{\varpi, tot} = \sqrt{ ( EIF. \, \sigma_\varpi ^\mathrm{DR3} ) ^2+(\sigma_\varpi ^{VLBI} )^2}}$, black line representing a standard Gaussian). The dashed lines correspond to the respective standard deviations of the Gaussians, while the solid straight lines show the mean values. The offset of the mean value with respect to 0 is a measure of the ZPO.}
\label{fig: Gaia calibration}
\end{figure*}

\begin{figure}[t]
\centering
\includegraphics[scale=0.4]{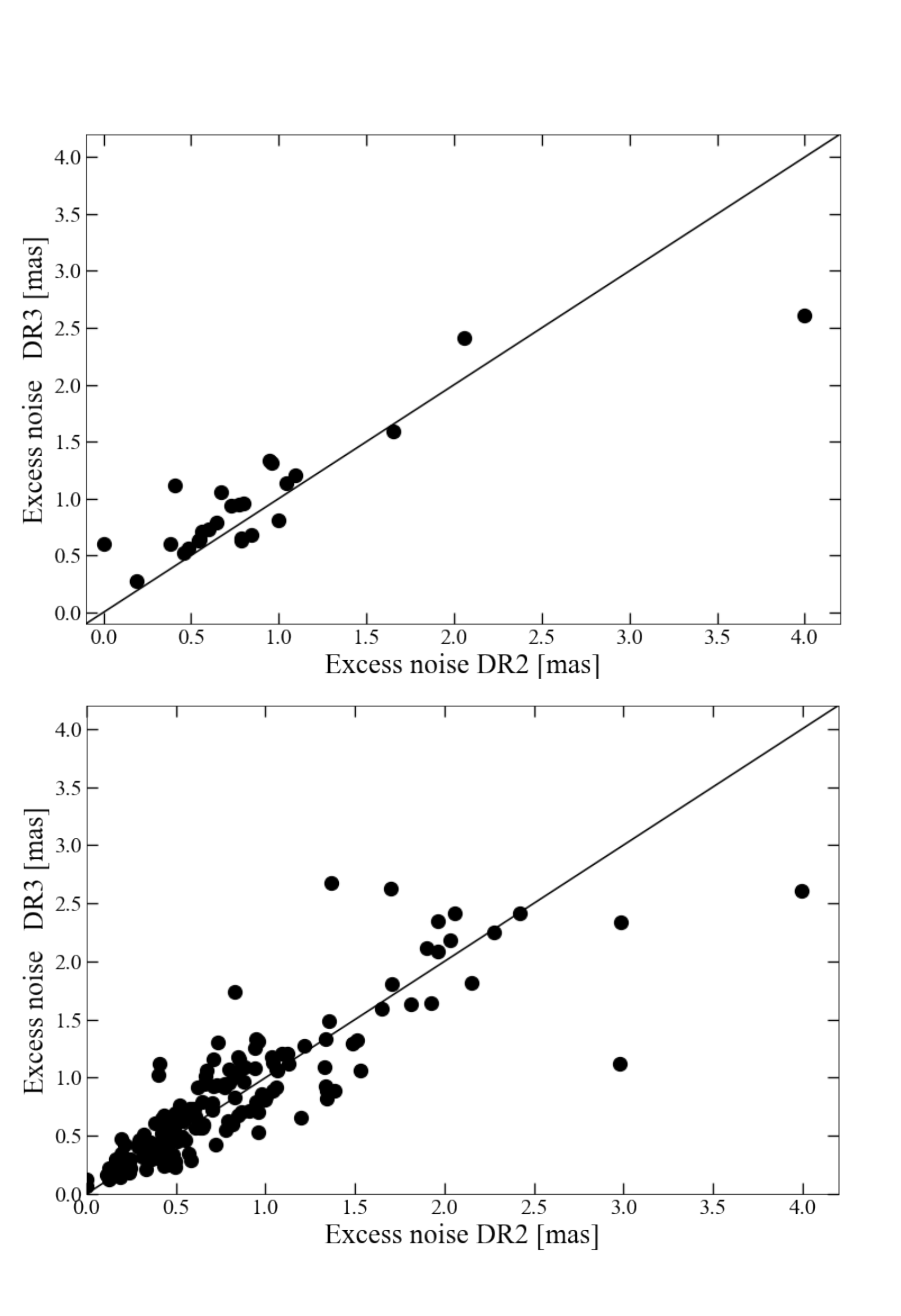}
 
\caption{Comparison between the astrometric excess noise of Gaia DR2 and DR3 for the VLBI sample (top) and the DEATHSTAR sample (bottom). The solid lines show the 1-to-1 relation. }
\label{fig: excess noise}
\end{figure}

In this work, we calibrated the \textit{Gaia} DR3 parallaxes for AGB stars using the sample of VLBI sources described in Sect.~\ref{sec: VLBI sample}, which is larger than, and includes, the sample of \citet{VanLan2018}. To that end, we normalised the distribution of the difference $\Delta  \varpi =  \varpi_\mathrm{ DR3} - \varpi_\mathrm{VLBI} $  by the sum of the uncertainties in quadrature. We then adjusted the \textit{Gaia} DR3 uncertainties until that normalised statistical distribution of the parallax difference had the properties of a standard Gaussian, with 0 mean and a standard deviation of 1, as illustrated by Fig.~\ref{fig: Gaia calibration}. This was done for the three categories described in Sect.~\ref{sec:Gmag color}. The correction factor to be applied to the \textit{Gaia} DR3 parallax errors or the error inflation factor ($EIF$), that is,
\begin{equation}
    \sigma_{\varpi,\mathrm{tot}} ^\mathrm{cor}= \sqrt{ ( EIF. \, \sigma_\varpi ^\mathrm{DR3} ) ^2+(\sigma_\varpi ^{VLBI} )^2}\, \mathrm{mas}\, ,
\end{equation}
and the zero-point offset ($ZPO$) of the parallax,
\begin{equation}
    \varpi_\mathrm{DR3} ^\mathrm{cor}=  \varpi_\mathrm{DR3} - ZPO\,\mathrm{mas}\, ,
\end{equation}
are given in Table~\ref{table: EIF and ZPO}.
Figure~\ref{fig: Gaia calibration} also shows the cases where no correction was applied to the total error, that is, ${\sigma_{\varpi, \mathrm{tot}} = \sqrt{ (\sigma_\varpi ^\mathrm{DR3}) \, ^2+(\sigma_\varpi ^{\mathrm{VLBI}})\, ^2}}$~mas, and where the astrometric excess noise was added to the \textit{Gaia} standard errors, ${\sigma_{\varpi, \mathrm{tot}}= \sqrt{(\sigma_\varpi ^\mathrm{DR3}) ^2+\sigma^2 _\mathrm{\small{excess\,noise}} \, + (\sigma_\varpi ^\mathrm{VLBI}) \,^2}}$~mas, as in \citet{VanLan2018}. Since the distribution of the parallax difference shown in Fig.~\ref{fig: Gaia calibration} was normalised by the total error, only taking into account the standard \textit{Gaia} DR3 uncertainties without any correction led to an underestimation of the total error, resulting in a parallax difference distribution broader than the standard Gaussian for the sources brighter than $G=12$~mag. However, including the astrometric excess noise parameter to the total uncertainty of the \textit{Gaia} DR3 parallax of these bright sources overestimated the total error, which led to a narrower parallax difference distribution ($\Delta  \varpi/\sigma_ {\varpi, \mathrm{tot}}$), as indicated by the blue line in Fig.~\ref{fig: VLBI vs Gaia}. Furthermore, Fig.~\ref{fig: excess noise} shows that the astrometric excess noise parameter for \textit{Gaia} DR3 is larger than for the DR2 parallaxes for the VLBI sources. The median value of the ratio of astrometric excess noises DR3/DR2 for the VLBI sample is 1.22, and 76\,\% of the sources have an astrometric excess noise DR3/DR2 ratio larger than 1. For the DEATHSTAR sample, close to 60\,\% of the sources have an astrometric excess noise higher in DR3 than in DR2. \\
Our results show that the parallax errors of the brightest sources ($G<8$) are dramatically underestimated, by more than a factor of 5. For sources between 8 and 12 $G$~mag, the nominal errors are larger, so a relatively smaller correction was needed. For the faintest stars, the nominal \textit{Gaia} DR3 parallax errors are, in principle, already very large, so little correction was needed on the \textit{Gaia} DR3 parallax errors. For some of these faint sources, the \textit{Gaia} DR3 parallax errors are so large that they would have needed to be reduced to recover the corresponding VLBI parallaxes.  Distances obtained with parallaxes with such large errors are likely to be very uncertain (see Sect.~\ref{sec: Bayesian distance}). Only 3 VLBI sources are fainter than $18$~mag, which is a too small sample to obtain statistically significant results. Therefore, we did not apply any correction to the faintest stars ($G \geq 12$~mag). When considering the VLBI sample as a whole (33 stars), we obtained a constant $EIF$ of $\sim4$. However, it is clear from Figs. \ref{fig: VLBI Gaia parallax}--\ref{fig: DEATHSTAR Gaia parallax} and Table \ref{table: EIF and ZPO} that not all sources require the same correction factor, so a constant $EIF$ is not suitable.

Using the properties of resolved binaries to calibrate the \textit{Gaia} eDR3 parallaxes, \citet{El-Badry2021} found that the published \textit{Gaia} eDR3 parallax uncertainties are underestimated by $\sim\,$30 to 80\,\% for bright red sources ($G<12$~mag), but are generally more reliable for fainter sources ($G\geq18$~mag), which is in agreement with our findings on AGB stars. The dependence of the parallax $ZPO$ of the \textit{Gaia} eDR3 parallaxes on the sky position and magnitude are discussed in detail in \citet{Lindegren2021} and \citet{Martin2021}, for instance, using samples of quasi-stellar objects and wide binaries. The zero-point offsets that we obtained for AGB stars, listed in Table \ref{table: EIF and ZPO}, are much larger, in absolute value, than the global eDR3 zero-point offset of $-0.028\,$mas derived by \citet{Ren2021}, and of $-0.039\,$mas by \citet{Martin2021}.
\begin{table}
\centering
 
\caption{\textit{Gaia} DR3 parallax zero-point offset ($ZPO$) and error inflation factor ($EIF$).}
    \begin{tabular}{cccc}
    \toprule
     $G$& $N_\mathrm{VLBI}$& $ZPO$ & $EIF$ \\
    \midrule
  $<8$ & $17$ &  $-0.077\pm0.004$ & $5.44\pm0.11$\\
  \addlinespace
  $\geq 8$ \& $<12$ & $13$ &  $-0.174\pm 0.006$ & $2.74^{+0.04} _{-0.05}$\\
  \addlinespace
  all$^{**}$ & $33$ &  $-0.131^{+0.016} _{-0.015}$ & $4.20 ^{+0.10} _{-0.11}$\\
    \bottomrule
    \multicolumn{3}{l}{\footnotesize  $N_\mathrm{VLBI}$: number of VLBI sources }
    \\
    \multicolumn{3}{l}{\footnotesize $^{**}$ all $G$ included, constant $ZPO$ and $EIF$.}
    \\
    
    \end{tabular}
    \label{table: EIF and ZPO}
\end{table}

\section{From parallax to distance}
\label{sec: Bayesian distance}
The distance $r$ of an object is equal to the inverse of its true parallax $ \varpi_\mathrm{True}$. Assuming that the measured parallax $ \varpi$ is a value taken from a normal distribution around the true parallax ${\varpi_\mathrm{True} = 1/r}$, with a known standard deviation $\sigma_ \varpi$, the likelihood of that measured parallax is given by
\begin{equation}
    P\,(\varpi\, \vert\, r, \sigma_\varpi) = \frac{1}{\sqrt{2 \pi}\, \sigma_\varpi} \, \mathrm{exp} \Big[ - \frac{1}{2\,\sigma_\varpi ^2} \Big( \varpi - \frac{1}{r} \Big)^2 \Big] \mathrm{,}
\end{equation}
where $\sigma_\varpi >0 $. \citet{BJ2015} showed that determining the distance from a probability distribution over the parallax presents several problems including negative parallaxes, in particular when the fractional error on the parallax is large \citep{Luri2018}.
\\A better approach is to infer the most probable value of the distance amongst all its possible values from the noisy measured parallax \citep{BJ2015, BJ2016, BJ2016b, BJ2018, BJ2021}. Such a posterior probability distribution over the distance is the product of the likelihood of the measurement, $P\,(\varpi\, \vert\, r, \sigma_\varpi)$, and an appropriate prior information on the distance, $P(r)$. Following Bayes theorem,
\begin{equation}
    P\,( r \, \vert\,\varpi, \sigma_\varpi) = \frac{1}{Z} P\,(\varpi\, \vert\, r, \sigma_\varpi) \,P(r) \, \mathrm{,}
\end{equation}
where $Z$ is a normalisation constant independent of the distance,
\begin{equation}
    Z= \int^{r_\mathrm{max}} _0 P\,(\varpi\, \vert\, r, \sigma_\varpi) \,P(r) \, \mathrm{d}r \, \mathrm{.}
\end{equation}
The value of the most probable distance is given by the median of the posterior distribution \citep{BJ2021}. The resulting distance uncertainties are asymmetrical because of the non-linear nature of the $1/\varpi$ to $r$ transformation.
\\
The determining factor in this distance inference problem is the fractional error on the parallax. For a parallax fractional error higher than 20\,\%, the posterior over the distance is highly asymmetrical \citep{BJ2015}. This effect worsens with increasing fractional error which results in a non-negligible increase in the value of the derived distance and its uncertainties, and therefore leads to less reliable distances. About 63\,\% of the sources published in eDR3 have parallax fractional errors larger than 20\,\% \citep{BJ2021}.\\

We calculated the posterior distribution over the distance for the sources in the VLBI and the DEATHSTAR samples. In that process, the choice of the prior was paramount. The more assumptions are introduced into the prior, the more model-dependent the resulting distances are, especially when the errors involved are large. Therefore one should choose a prior that closely represents the expected data. To this end, we tested four different priors:
a uniform distribution (UD) prior defined in
\citet{BJ2015} as
    \begin{equation}
        P_\mathrm{UD}=\begin{cases}
         \frac{1}{r_\mathrm{max}}, & \text{if $0<r<r_\mathrm{max}$}\\
        0, & \text{otherwise}
        \end{cases}
        \mathrm{;}
    \end{equation}
a uniform space distribution (USD) prior that is corrected for increasing volume, following the definition given by \cite{BJ2015}
     \begin{equation}
        P_\mathrm{USD}= \begin{cases}
       {\huge \frac{3\,r^2}{r_\mathrm{max} ^3}}, & \text{if $0<r<r_\mathrm{max}$}\\
        0, & \text{otherwise}.
        \end{cases}
        \mathrm{;}
    \end{equation}
 an exponentially decreasing space distribution (EDSD) prior that exponentially decreases to 0 as the distance increases, given by
    \begin{equation}
        P_\mathrm{EDSD}=\begin{cases}
        \frac{r^2}{2L^3} \, \mathrm{e}^{-(r/L)}, & \text{if $r>0$}\\
        0, & \text{otherwise}
        \end{cases}
    \end{equation}
    \citep{BJ2015}, with $L=250\,$pc; and
a more realistic prior that describes the Galactic distribution of AGB stars (AGB prior). According to \citet{Jura1990,Jura1992_tenaizy}, the distribution of AGB stars in the Galaxy follows a vertical scale height up to $Z_0=240\,\mathrm{pc}$, and a disc scale length set at $R_0$ of $3500\,\mathrm{pc}$. The AGB prior is also corrected for increasing volume, so that
    \begin{equation}
        P_\mathrm{AGB}=\begin{cases}
         \frac{3\,r^2}{r_\mathrm{max} ^3} \, \, \mathrm{e}^{-(|z| /Z_0)} \, \mathrm{e}^{-(R_\mathrm{Gal}/R_0)}, & \text{if $0<r<r_\mathrm{max}$}\\
        0, & \text{otherwise}
        \end{cases}
        \mathrm{,}
    \end{equation}
    where $R_\mathrm{Gal}$ is the distance to the Galactic centre and $z$ the height above the Galactic plane.
    
By checking the 2MASS $K$-band magnitudes of our sample, we found an apparent $K$-band limit of 3.2 mag. Assuming that the tip of the AGB is at $V=-2$ for a solar-mass star, and taking an average of $(V-K)=8$ from theoretical estimates \citep{Sara2013}, we obtained a maximum distance estimate of $4400\,\mathrm{pc}$. We therefore limited the allowed distances $r_\mathrm{max}$ for the sources in our sample to $r_\mathrm{max}=4500 \, \mathrm{pc}$.\\

The Galactic distribution that defines the AGB prior used in this work, based on the findings of \citet{Jura1992_tenaizy}, is in agreement with the Galactic distribution derived by \citet{Ishihara2011} who investigated the difference between the distribution of AGB stars in the Milky Way for the M- and C-type stars. They found a concentration of oxygen-rich stars toward the Galactic centre, with a density decreasing with Galactocentric distance, and a rather uniform distribution within about 8$\,$kpc of the Sun for carbon-rich AGB stars. \citet{Tom2002} found a distribution that follows a vertical scale height up to 300$\,$pc and a constant number of AGB stars in the radial direction, up to 5 kpc, above which the density decreases exponentially with a scale length of 1.6 kpc, extending to $\sim 12\,$kpc. The scale length that we adopted in the AGB prior is consistent with the results of both studies. Varying the scale height $Z_0$ or scale length $R_0$ did not significantly change the derived distances (less than 10\,\% of deviation). The only parameter that affected the derived distances, mostly for the farthest stars, was the maximum allowed distance $r_\mathrm{max}$. Lowering its value led to an increase in the number of sources whose distances were stuck at that upper limit because the posteriors could not reach convergence. Increasing $r_\mathrm{max}$ to 5000$\,$pc only changed the distance of two of the stars in the DEATHSTAR sample, putting them at that new upper limit, implying that the distances to these sources are very poorly constrained.\\
\begin{figure*}[t]
\centering
\includegraphics[scale=0.23]{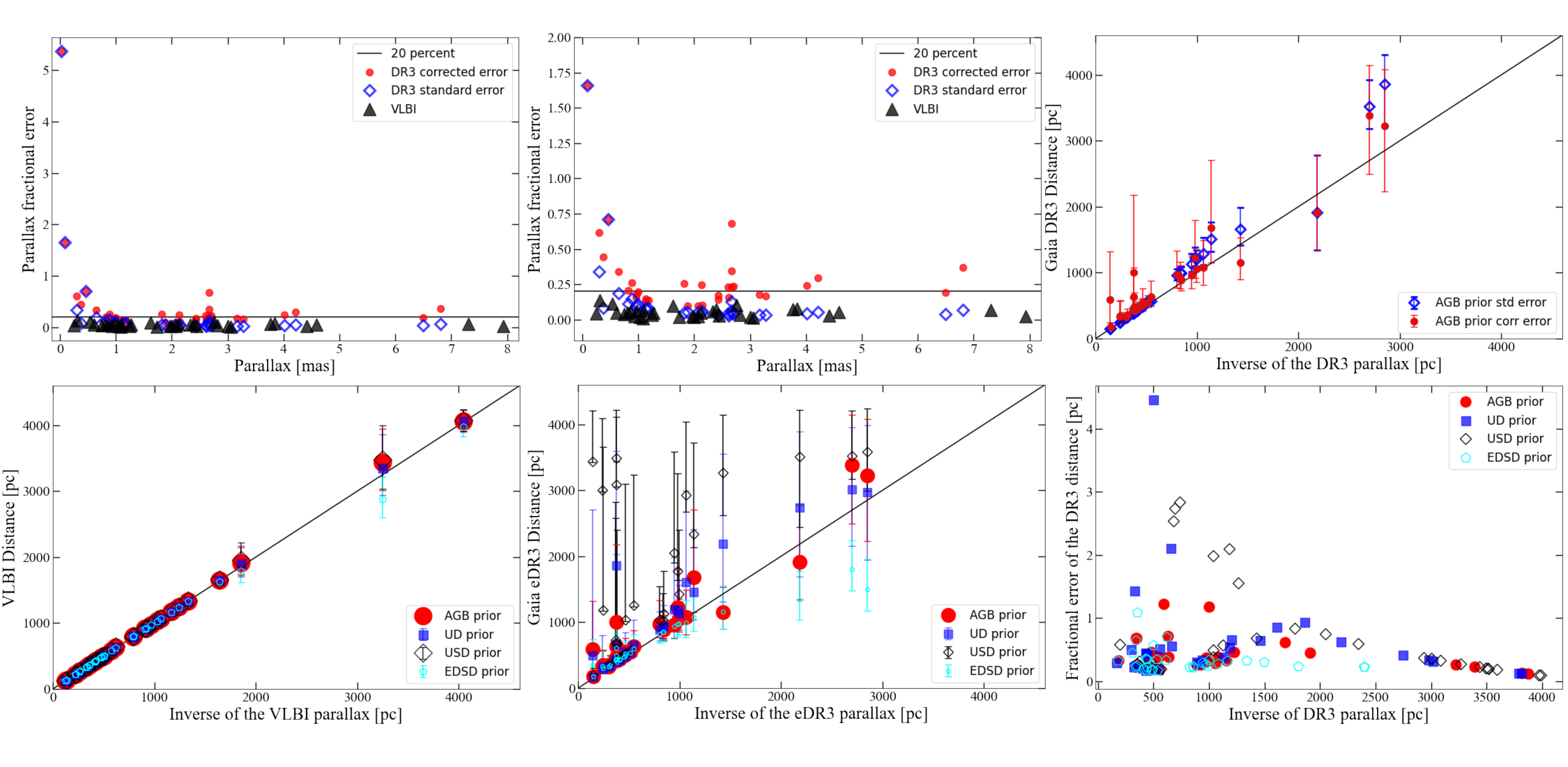}
 
\caption{\textit{Gaia} DR3 and VLBI parallaxes with the corresponding distances for the VLBI sample. \textbf{Top panels:} Fractional errors of the VLBI (black full triangles) and \textit{Gaia} DR3 parallaxes using the standard nominal error (blue open diamond) and with the corrected errors (red full circles) for the VLBI sources. The plot in the middle is a zoomed version of the plot on its left. The plot on the right compares the distances derived with \textit{Gaia} DR3 parallaxes with the corrected (red) and the standard errors (blue). \textbf{Bottom panels:} Distances derived using the four priors for the VLBI sources using the VLBI parallaxes (left) and the corrected \textit{Gaia} DR3 parallaxes (middle). The solid lines represent the 1-to-1 relation. The plot at the bottom-right shows the fractional error on the derived distances.  }
\label{fig: VERA distance}
\end{figure*}
\begin{figure*}[t]
\centering
\includegraphics[scale=0.24]{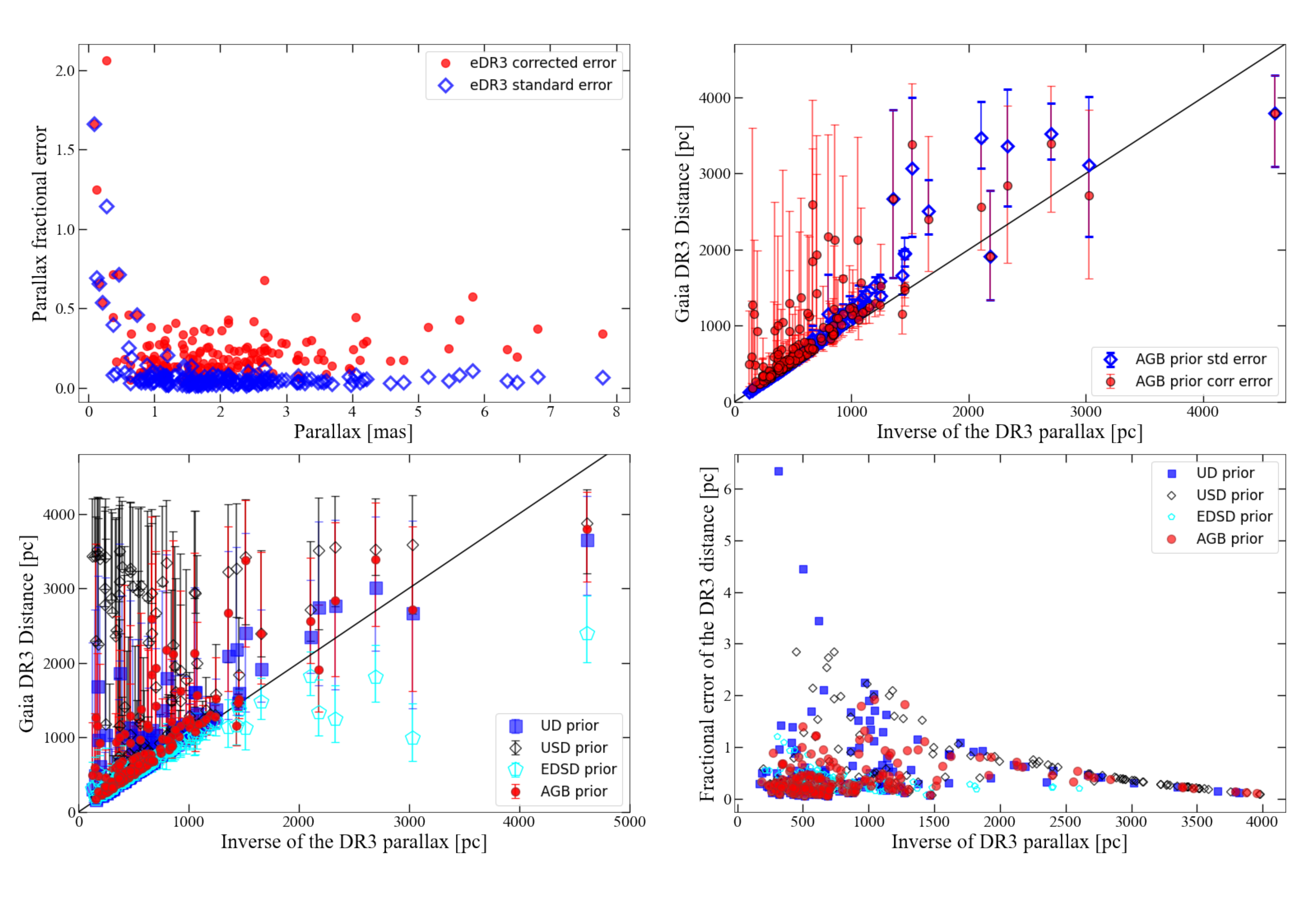}
 
\caption{\textit{Gaia} DR3 parallaxes and the corresponding distances for the DEATHSTAR sample. \textbf{Top panels:} Fractional errors of the \textit{Gaia} DR3 parallaxes using the standard nominal error (blue open diamond) and with the corrected errors (red full circles) for the DEATHSTAR sources. The right plot compares the distances derived with \textit{Gaia} DR3 parallaxes with the corrected (red) and the standard errors (blue). \textbf{Bottom panels:} Distances derived with the four priors for the DEATHSTAR sources using the corrected \textit{Gaia} DR3 parallaxes. The solid lines represent the 1-to-1 relation. The plot on the right shows the fractional error on the corrected \textit{Gaia} DR3 distances.}
\label{fig: DEATHSTAR distance}
\end{figure*}

We find that parallax fractional errors larger than $\sim$ 18\,\% already lead to notable errors in the derived distances ($\gtrsim 20$\,\% distance error with the AGB prior) for some of the sources. More significant uncertainties are associated with parallax fractional errors greater than the previously mentioned $20$\,\% cutoff. About 85\,\% of the VLBI sample have a \textit{Gaia} DR3 parallax fractional error smaller than 0.2 when considering the standard error, but that number decreases to $\sim$\,35\,\% after applying the correction to the \textit{Gaia} DR3 parallaxes derived in Sect.~\ref{sec: correction}. In the case of the DEATHSTAR sample, 94\,\% of the sources have a fractional parallax error lower than $20$\,\% before correction, which decreases to $\sim$\,52\,\% after correcting the errors on the DR3 parallaxes (see Table~\ref{table: dist Append1}).

As expected, the distances derived with the VLBI parallaxes are mostly unchanged with the four priors due to the low measurement uncertainties, as illustrated in Fig.~\ref{fig: VERA distance}. The fractional errors on the VLBI parallaxes are lower than 20\,\% for almost all the sources. In such cases, the posterior distribution is dominated by the likelihood of the data, and the choice of prior does not change the derived distances. The uncertainties on the derived distances are also mostly smaller than 20\,\%, showing their high level of reliability.
\\
The distances derived with the corrected \textit{Gaia} DR3 parallaxes
are strongly dependent on the prior, as the latter dominates the posterior at large parallax fractional errors. The effects of the value of the parallax fractional errors on the derived distances are illustrated in Figs.~\ref{fig: VERA distance} and \ref{fig: DEATHSTAR distance}, showing comparisons between the \textit{Gaia} DR3 distances obtained with the different priors, and using the standard and the corrected errors of the \textit{Gaia} DR3 parallaxes for the sources in the VLBI (Fig.~\ref{fig: VERA distance}) and the DEATHSTAR samples (Fig.~\ref{fig: DEATHSTAR distance}).

In the following, we focus on the distances estimated using the AGB prior, as it is more informative and realistic. It is also more sensitive than the other priors we tested because of the larger number of parameters in it.
The larger the uncertainties, the higher the probability for the true parallax to be smaller than the measured parallax. As a result, our estimate of the true distance increases. This is known as the Lutz–Kelker bias \citep{Lutz1973}. The fractional error on the measurements and the sensitivity of the prior determine how bad the effects are, which means that even nearby objects can be affected. The large errors on the parallaxes lead to significant  errors on the derived distances, making them unreliable. Using the corrected \textit{Gaia} DR3 parallaxes and the AGB prior, we found that about 46\,\% of the sources in the DEATHSTAR sample have distance fractional errors larger than 25\,\%, while more than 15\,\% of them have distance fractional errors greater than $50$\,\%. When the prior and/or the posterior distribution did not converge or when the uncertainty on the derived distance was too large, we rejected the derived distance.

\section{A new PL relation for Miras}
\label{sec: PL}
A number of the distances derived with the corrected \textit{Gaia} DR3 parallaxes following the method described in Sect.~\ref{sec: Bayesian distance} are highly uncertain for the sources in the DEATHSTAR sample, and thus rejected. Therefore, we turned to a different method to estimate the distances to these sources: the PL relation. We derived a new bolometric PL relation based on the VLBI distances of the Mira variables in the VLBI sample.

The luminosity of the sources in the VLBI sample were obtained by modelling their spectral energy distributions (SEDs) using the radiative transfer code DUSTY \citep{Ivezic1999}. DUSTY solves the radiative transfer through a dusty environment including dust absorption, emission, and scattering. The details of the SED fitting are given in Appendix~\ref{sec: Append dusty}. From the dust modelling, we obtained the effective temperature of the central star, $T_{\star}$, the dust temperature at the inner radius, $T_{\rm{d}}$, the optical depth at 10 $\mu$m, $\tau_{10}$, and the bolometric luminosity, $L_{\star}$, of each source in the VLBI sample. These results are listed in Table~\ref{table: RT}.

{
\renewcommand{\arraystretch}{1.07}
\begin{table*}
\centering
 
\caption{DUSTY results for the VLBI sources.}
    \begin{tabular}{lcccccccccc}
    \toprule
     Source & $r_\mathrm{VLBI}$ & $\sigma_{r, \mathrm{VLBI}} ^-$ & $\sigma_{r, \mathrm{VLBI}} ^+$& $L_\star ^\mathrm{median}\,$&$\sigma_{L_\star} ^\mathrm{-}\,$& $\sigma_{L_\star} ^\mathrm{+}\,$ & $T_\star$& $T_\mathrm{dust}\,$& $\tau_{10}$ & PL$^*$\\
     & [pc] & [pc] &[pc] & [L$_\odot$]&[L$_\odot$]&[L$_\odot$] & [K] & [K] \\
      \midrule
      \textit{Miras} \\
AP Lyn & 501 & 10 & 10 & 4200 & 100 & 200 & 2900 & 700 & 0.28 & no\\
BX Cam & 579 & 10 & 10 & 7700 & 200 & 200 & 3000 & 1000 & 0.8 & yes\\
FV Boo & 1034 & 60 & 67 & 1900 & 200 & 200 & 3000 & 700 & 0.3 & no\\
NSV 17351 & 4064 & 157 & 168 & 25600 & 1400 & 1600 & 3100 & 1000 & 1.2& no \\
OZ Gem & 1246 & 58 & 63 & 2500 & 100 & 200 & 2700 & 600 & 0.7 & no\\
QX Pup & 1652 & 78 & 86 & 1300 & 100 & 100 & 3600 & 600 & 5.0& no \\
R Aqr & 220 & 11 & 12 & 8100 & 600 & 600 & 2900 & 1200 & 0.14& yes \\
R Cnc & 266 & 19 & 22 & 4800 & 500 & 600 & 2900 & 600 & 0.02& yes \\
R Hya & 126 & 2 & 3 & 10300 & 300 & 300 & 3100 & 1200 & 0.04& yes \\
R Peg & 374 & 36 & 44 & 4600 & 600 & 700 & 2900 & 1200 & 0.11 & yes\\
R UMa & 508 & 12 & 14 & 4100 & 100 & 200 & 3000 & 1200 & 0.17 & yes\\
RR Aql & 411 & 11 & 12 & 3500 & 100 & 200 & 2900 & 900 & 0.26 & yes\\
S CrB & 424 & 28 & 33 & 8600 & 900 & 800 & 2900 & 1000 & 0.07 & yes\\
S Ser & 801 & 25 & 27 & 5800 & 300 & 300 & 3000 & 600 & 0.04 & yes\\
SY Aql & 922 & 56 & 64 & 2800 & 200 & 300 & 3000 & 700 & 0.24& no \\
SY Scl & 1330 & 50 & 55 & 5700 & 300 & 300 & 3000 & 800 & 0.18& yes \\
U Her & 271 & 19 & 21 & 5800 & 600 & 600 & 2900 & 1000 & 0.08 & yes\\
T Lep & 327 & 4 & 4 & 6100 & 200 & 100 & 2900 & 1200 & 0.06 & yes\\
U Lyn & 792 & 36 & 39 & 6000 & 400 & 400 & 3000 & 900 & 0.16& yes \\
UX Cyg & 1918 & 198 & 250 & 4000 & 600 & 700 & 3600 & 600 & 5.0& no \\
V837 Per & 918 & 9 & 8 & 5400 & 100 & 100 & 3100 & 1200 & 1.0& yes \\
W Leo & 971 & 18 & 19 & 6800 & 200 & 200 & 2600 & 600 & 0.04 & yes\\
X Hya & 484 & 11 & 12 & 3800 & 100 & 100 & 3000 & 900 & 0.09 & yes\\
Y Lib & 1173 & 64 & 73 & 3200 & 200 & 300 & 3000 & 700 & 0.06& yes \\
\\
\textit{SRa}\\
HU Pup & 3437 & 426 & 51 & 29950 & 3450 & 3450 & 2600 & 600 & 0.48 & no\\
RW Lep & 636 & 59 & 72 & 10300 & 1400 & 1500 & 3000 & 600 & 0.02 & no\\
\\
\textit{SRb and U}\\
BX Eri & 476 & 23 & 25 & 6500 & 400 & 500 & 3100 & 600 & 0.03& no \\
HS UMa & 356 & 11 & 13 & 6100 & 300 & 300 & 3100 & 600 & 0.03& no \\
RT Vir & 227 & 6 & 7 & 7400 & 300 & 300 & 3100 & 600 & 0.02& no \\
RX Boo & 139 & 9 & 11 & 4700 & 500 & 600 & 3100 & 600 & 0.02& no \\
S Crt & 433 & 23 & 25 & 4800 & 400 & 400 & 3200 & 600 & 0.01& no \\
SV Peg & 334 & 7 & 7 & 8700 & 300 & 400 & 3100 & 600 & 0.02 & no\\
V637 Per & 1065 & 22 & 23 & 4500 & 200 & 200 & 3100 & 600 & 0.04& no\\
    \bottomrule
    \multicolumn{9}{l}{\footnotesize SRa/b: semi-regular a or b., U: unknown} \\
    \multicolumn{9}{l}{\footnotesize $r_\mathrm{VLBI}$ the median distance obtained with the VLBI parallax} \\
    \multicolumn{9}{l}{\footnotesize $\sigma_\mathrm{r,VLBI} ^\mathrm{-/+}$ the lower/upper uncertainties on the VLBI distance.} \\
    \multicolumn{9}{l}{\footnotesize $L_\star ^\mathrm{median}$ the median luminosity; $\sigma_{L_\star} ^\mathrm{-/+}$ the lower/upper uncertainties on the luminosity.} \\
    \multicolumn{9}{l}{\footnotesize $T_\star$ is the stellar temperature; $T_\mathrm{dust}$ the dust temperature; $\tau_{10}$ the optical depth at 10 $\mu$m;} \\
    \multicolumn{7}{l}{\footnotesize $^*$included in the PL derivation.} \\
    \end{tabular}
    \label{table: RT}
\end{table*}
}
The bolometric magnitude $M_\mathrm{bol}$ of each star was calculated (assuming $M_{\rm{bol},\odot}$=4.74) and correlated with the variability period which was taken either from the GCVS \citep{GCVS} or the International Variable Star Index \citep[VSX; ][]{Watson2021}. We removed from the PL relation derivation the sources that are known or suspected not to behave as regular Mira variables.
The OH/IR stars in the VLBI sample were, therefore, excluded from this analysis. These stars lie at the end of the evolution of the AGB and have very high mass-loss rates \citep{Herman1985}. They have been found to consistently lie below the PL relation of typical Miras \citep{vanlang1990, Whitelock1991}.  The stars in our VLBI sample labelled as OH/IR in the SIMBAD database are NSV~17351, QX~Pup, and SY~Aql. In addition, OZ~Gem has recently been suspected as being in the process of becoming an OH/IR star \citep{Urago2020}. The study conducted by \citet{Chibueze2020} suggests that AP~Lyn could undergo the same process as OZ~Gem due to its high mass-loss rate. Finally, we also excluded FV~Boo and UX~Cyg from the Mira-PL relation determination. \citet{Lewis2002b,Lewis2002} described FV~Boo as a dying OH/IR star, with a mass loss that abruptly switches on and off. Monitoring of FV Boo in the IR over a 20-year span by \citet{Kamezaki2016fvboo} showed that it suffered a temporary but significant dip in its luminosity in 2005. The star UX~Cyg stood out in the sample of Mira variables investigated by \citet{Etoka2000} due to rapid, large changes in the amplitude of the variations of its maser lines, likely linked to an unusually turbulent envelope. \citet{White2008} excluded UX~Cyg from their $K$-band PL relation as they hypothesised it to be either an overtone pulsator or a hot bottom burning candidate.\\

We used the form of the PL relation introduced by \citet{White2008} where the zero-point is shifted within the range of typical periods of Mira variables. The resulting PL relation has a slope of $-2.67\pm1.88$ and a zero-point of $2.323\pm4.832$.
The large errors are due to the small sample size. This problem can be solved by using a fixed slope and by keeping the zero point as only free parameter.
Using stars in the Large Magellanic Cloud (LMC) and in the Milky Way, \citet{White2008} showed that, while the zero-point differs for oxygen-rich and carbon stars, the slope of the $K$-band PL relation is nearly invariant. In addition, using a common fixed slope for the Milky Way and the LMC, \citet{White2008} showed that the zero-point of the $K$-band PL relation for Mira variables in these two different environments were consistent with each other within their uncertainties, implying that the $K$-band PL relation is universal. This property of universality was then extended to the bolometric PL relation by \citet{White2009}, where they determined the distance to the Fornax Galaxy using a bolometric PL relation derived in the LMC. Accordingly, with the assumption that the slope invariance with chemical type is also applicable to the bolometric PL relation, we use the slope of $\rho_\mathrm{fixed}^\mathrm{shift}=-3.31\pm0.24$ derived for C-stars in the LMC by \citet{White2009}. We obtain the PL relation given by
\begin{equation}
    M_\mathrm{bol} = (-3.31\pm0.24)\,[\mathrm{log}\, P - 2.5] + (-4.317\pm0.060)\mathrm{.}
    \label{eq: PL}
\end{equation}
\begin{figure}[t]
\centering
\includegraphics[scale=0.42]{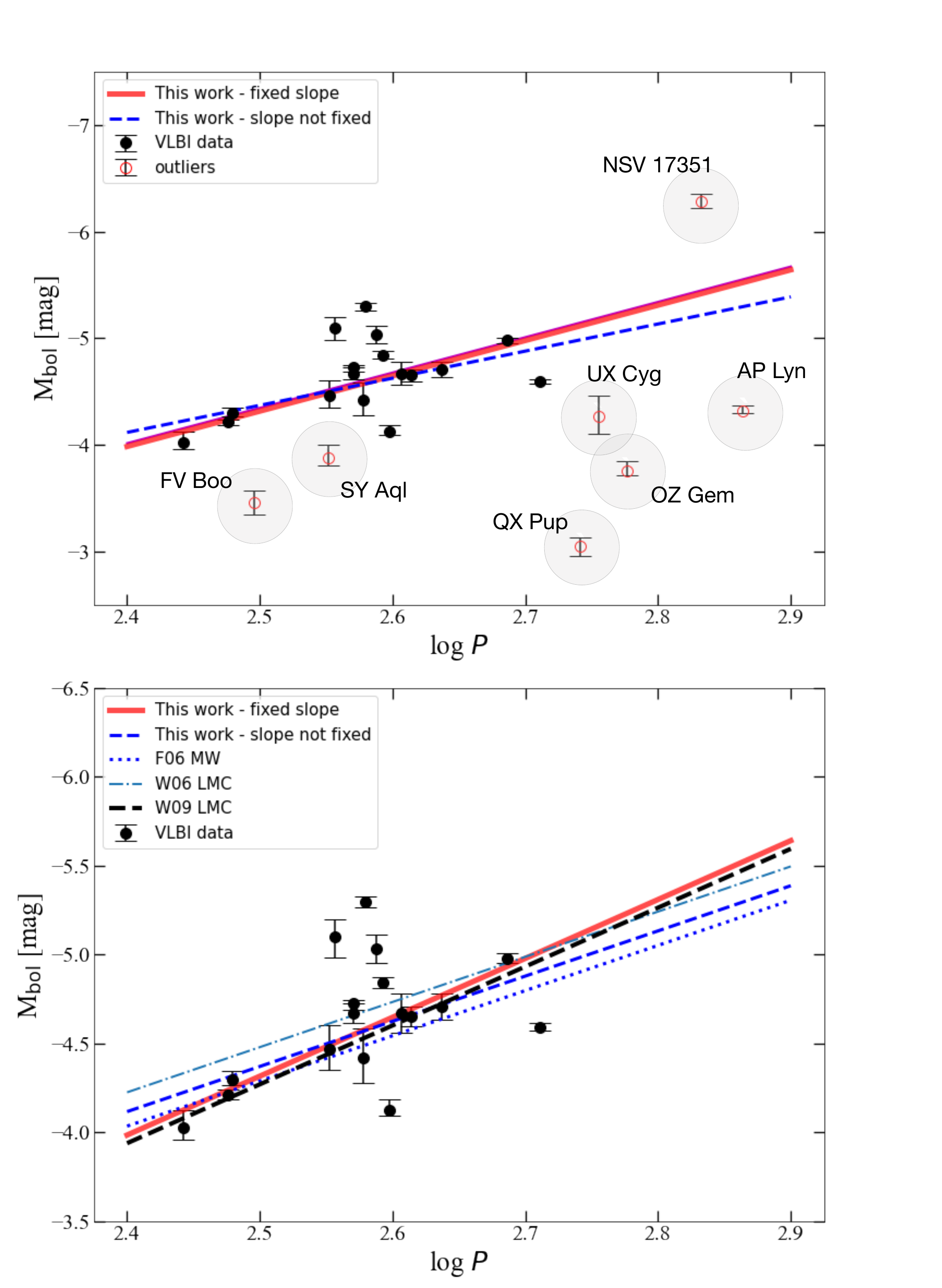}
 
\caption{Derived PL relation for Mira variables. \textbf{Top}: Bolometric PL relation based on the luminosity of the Mira variables in the VLBI sample. The red solid line shows the PL relation obtained when using a fixed slope, and the blue dashed line represents the relation obtained when the slope is a free parameter. The sources excluded from the fit are represented by the open symbols. \textbf{Bottom}: Comparison between our PL relation with the bolometric PL relations for C-type stars by \citet{feast2006} in the Milky Way (F06 MW), \citet{White2006} (W06 LMC), and \citet{White2009} (W09 LMC) in the LMC.}
\label{fig: PL}
\end{figure}
The zero-point derived in this work for M-type stars reasonably agrees with existing bolometric PL relations for C-stars both in the LMC \citep{White2006, White2009} and in the Milky Way \citep{feast2006}, as seen in Fig.~\ref{fig: PL}. In particular, our PL relation is in excellent agreement with the zero-point of the PL relation of \citet{White2009}. \citet{White2008,White2009} compared the luminosity obtained from interpolating IR measurements with the luminosity derived from dust modelling using DUSTY \citep[e.g.][]{Matsuura2007} and concluded that the latter could be overestimated by up to 50\,\%. On the other hand, bolometric magnitudes in \citet{White2006, White2009} were estimated by integrating under a spline curve fitted to fluxes at $J, H, K, L, 12$, and $25\,\mu$m . The curve was extrapolated at the end points to reach zero flux at zero frequency at the short end (long end in wavelength) and by joining the $K$-band flux with zero flux through a point between the $H$ and $K$ fluxes, which could have underestimated the total luminosity. However, the bolometric magnitude obtained with the PL relation derived in this work is only about 0.046 mag brighter than in \citet{White2009}, and 0.048 mag fainter than the bolometric magnitudes obtained by \citet{White2006}, on average. Furthermore, although the associated uncertainties are very large  because of the small sample size, the PL relation we derived with a non-fixed slope appears, in general, to be consistent with the PL relation derived by \citet{White2006} (see Fig.~\ref{fig: PL}). As expected, most of the (candidate/dying) OH/IR stars lie below the derived PL relation, with the exception of NSV~17351.
\subsubsection*{The universality of the PL relation}
As previously mentioned, \citet{White2008} demonstrated that the PL relation is universal. This was proven by the good agreement between the zero-points of the PL relations in the  LMC, the Milky Way, and the Fornax dwarf Galaxy, assuming a common slope. However, recent studies by \citet{Urago2020} and \citet{Chibueze2020} have cast doubt on the universality of the PL relation in environments with different metallicities. Their conclusion of a non-universal PL relation comes from the position of the Galactic oxygen-rich star OZ~Gem on the LMC $K$-band PL diagram.
Given its luminosity, chemistry, and high mass-loss rate implied by its very red colour, \citet{Urago2020} concluded that OZ~Gem is likely an OH/IR star. Their photometric measurements, however, placed OZ Gem on the region for C-stars on the $K$-band PL relation in the LMC, below the line for M-type stars (see their fig. 9).
Our results, show that OZ Gem also lies below the bolometric PL relation for Mira variables in the Milky Way (see
Fig.~\ref{fig: PL}). This corroborates the results of \citet{Urago2020} on the OH/IR nature of this source, as OH/IR stars tend to have low luminosity for their relatively long period \citep[e.g. fig. 10 in][]{Whitelock1991} compared to typical Miras. They are also expected to be less luminous in the $K$-band because of dust obscuration from their thick CSE. In other words, OZ~Gem is an outlier even amongst the Mira variables in the Milky Way and its position in the ($K$-band or bolometric) PL relation does not represent the general trend for the typical Mira variables in the Galaxy. This interpretation, therefore, does not invalidate the universality of the PL relation for regular Miras.
However, the absence of typical oxygen-rich Miras at longer periods ($>600$ days) is apparent in Fig.~\ref{fig: PL}, as they are expected to be either brighter due to hot bottom burning, or fainter due to high mass-loss rates (e.g. OH/IR stars). There is, however, no such gap at longer periods in the $K$-band PL relation of oxygen-rich stars in the LMC. The apparent absence of regular oxygen-rich Miras at longer periods in the Milky Way could indicate that the universality of the PL relation with metallicity breaks down at long periods for oxygen-rich Miras.
\section{A new distance catalogue}
\label{sec: cat}
In this section, we present a new distance catalogue for the $\sim200$ sources in the DEATHSTAR sample, based on our results in Sects.~\ref{sec: correction}, \ref{sec: Bayesian distance}, and \ref{sec: PL}, and using alternative methods in the literature when applicable, as described below. Table~\ref{table: dist Gaia1} gives the distances and the associated uncertainties for the nearby AGB stars in the DEATHSTAR sample. The displayed distances are estimated using the following methods.\\
For the AGB stars in the VLBI sample, we estimated the distances and their errors using the VLBI parallaxes (Type = V in Table~\ref{table: dist Gaia1}), following the Bayesian approach using the AGB prior described in Sect.~\ref{sec: Bayesian distance}. The corresponding distances have fractional errors within 25\,\%. For the sources outside the VLBI sample that have a corrected \textit{Gaia} DR3 parallax fractional error below 15\,\%, the best distance estimate listed in Table~\ref{table: dist Gaia1} is the \textit{Gaia} DR3 distance obtained with the AGB prior (Type = G$_\mathrm{AGB}$). For sources with a corrected \textit{Gaia} DR3 parallax fractional error between 15 and 20\,\%, we checked if the fractional error on the distance derived using the Bayesian approach with the AGB prior is within 25\,\%. If that condition was fulfilled, the distance in Table~\ref{table: dist Gaia1} is the \textit{Gaia} DR3 distance with AGB prior (Type = G$_\mathrm{AGB}$). Otherwise, the distance was determined using a PL relation, provided that the source had a known period and variability type. The PL relation derived in Sect.~\ref{sec: PL},
        \begin{equation*}
              M_\mathrm{bol} = (-3.31\pm0.24)\,[\mathrm{log}\, P - 2.5] + (-4.317\pm0.060) \mathrm{\, ,}
        \end{equation*}
was used to derive the luminosity of the Mira variables (Type = PL(M) in Table~\ref{table: dist Gaia1}). We then performed radiative transfer modelling of the stellar and dust emission, as in Sect.~\ref{sec: PL}, to determine their distances. The details of the dust modelling are given in Appendix~\ref{sec: Append dusty}. It is important to note that this PL relation was derived using Mira variables with periods between 277 and 514 days, so the distances for the sources outside this range derived with this relation (Type = PL(M$_\mathrm{out}$) ) can be less reliable. However, given the good agreement between our PL relation and existing PL relations for Miras that are valid within a wider period range, $\sim 160$ - $930$\, days for \citet{feast2006}, for instance, using our PL relation for sources within that wider range of periods is a reasonable first approximation.
\\For semi-regulars (SRs), we used the PL relation from \citet{Knapp2003} given by
        \begin{equation*}
            M_K = -1.34(\pm 0.06)\,\mathrm{log} \, P -4.5(\pm0.35)
        \end{equation*}
to estimate the absolute magnitudes of the sources in the \textit{K}-band. The distance was obtained using the distance modulus relation (Type = PL(SRa/b) in Table~\ref{table: dist Gaia1}). The apparent magnitude in the \textit{K}-band was retrieved from the VSX  online search tool \citep{Watson2021}. Interstellar extinction in the $K$ band is expected to be low. For the sources in our sample that are also in \citet{Knapp2003} or \citet{White2008}, we used the $A_K$ coefficients calculated by \citet{Knapp2003} or the $A_V$ (lower limits) in \citet{White2008} for these individual sources to correct for reddening. Otherwise, we assumed a value of $A_K=0.02$~mag, representing the mean value of the $A_K$ distribution in \citet{Knapp2003}. The use of a single PL relation for SRs is motivated in the literature by the fact that they are possible progenitors of  Mira variables, so some sort of similarity in their behaviour is expected \citep{FeastWhite2000}. The PL relation for SRs by \citet{Knapp2003} that we used here is in good agreement with the relation derived by \citet{Ye2004} for SRs, within their respective uncertainties. However, as SRs, SRbs in particular, are less regular and have lower amplitude, their pulsation behaviour is less well-understood than Mira variables, and their period and luminosity show a weaker correlation. In addition, studies such as \citet{Soszy2013} showed that SRs can have more than one pulsation period and lie on different PL sequences. Therefore, the reliability of the distances derived with this method can be arguable. More recently, \citet{Trabucchi2021} investigated the suitability of SRs as distance indicators and found that a subgroup of SRs follows the same sequence as Mira variables in the PL diagram. However, they concluded that long-time series are necessary to properly classify SRs according to their pulsation periods.  The study of the variability of SRs is beyond the scope of this paper. Finally, for the remaining sources in the DEATHSTAR sample with corrected \textit{Gaia} DR3 parallax fractional error larger than 20\,\%, we estimated the distance with a PL relation, depending on the variability type, as described above.

We note that the distances presented in this catalogue were derived in a systematic way for the whole DEATHSTAR sample. For some sources, however, independent and more accurate source-specific distances obtained with the phase-lag method are available in the literature. This is the case, for example, for IRC+10216 \citep[123$\pm14$ pc;][]{Martin2012}, R Scl \citep[361$\pm44$ pc;][]{Matthias2018}, and OH/IR stars \citep[e.g.][]{Herman1985,Engels2015, etoka2018}. Our distance of 408$^{+42} _{-34}$~pc for R~Scl agrees rather well with its phase-lag distance. The distance of 190$\pm{20}$~pc that we derived for IRC+10216 with our PL relation for Miras is consistent within 3\,$\sigma$ with its distance derived by \citet{Martin2012}, but the error is, nonetheless, relatively large. This can be due to the fact that the period of IRC+10216 lies outside the period range with which our PL relation was derived. This shows that one should be careful when using the PL(M$_\mathrm{out}$) distances in Table~\ref{table: dist Gaia1}, as the displayed errors do not account for uncertainties related to the longer/shorter periods of these sources.
\subsection*{Comparing the corrected \textit{Gaia} DR3 and the PL distances}
\begin{figure}
   \includegraphics[scale=0.27]{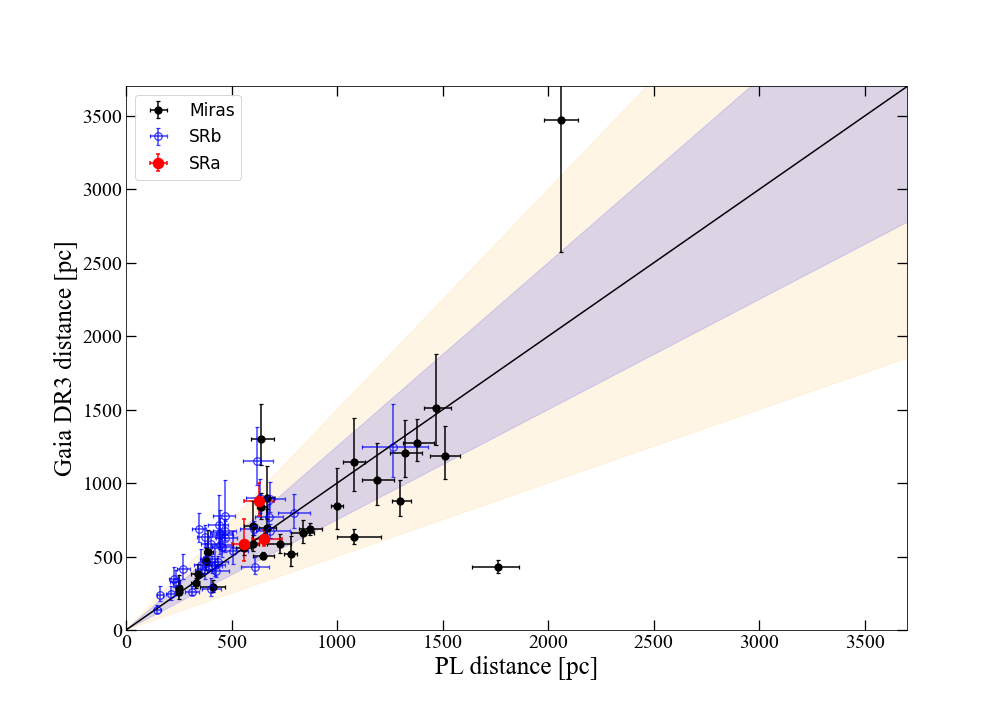}
    
    \caption{Comparison between the 'good' \textit{Gaia} DR3 distances (see text) and the PL distances for the same sources. The black line shows the 1-to-1 relation. The blue and orange regions show the range of $\pm25$ and $\pm50$\,\%, respectively.}
    \label{fig:comparison Gaia PL}
\end{figure}

We considered the sources that have 'good' \textit{Gaia} DR3 distances, which are distances with fractional errors below 25\,\%. For comparison purposes, we also derived alternative distances for these same sources, using either the PL relation that we developed in Sect.~\ref{sec: PL} for the Miras, or the PL relation  by \citet{Knapp2003} for the SRs. Figure~\ref{fig:comparison Gaia PL} shows that, for the Miras, the distances derived with the two methods are in good agreement, considering their respective uncertainties. The only notable exception is the carbon star LP~And, whose derived PL distance is about 4 times larger than its \textit{Gaia} DR3 distance. The nominal \textit{Gaia} DR3 parallax fractional error of LP~And is about 10~\%, but the corresponding astrometric excess noise is almost as large as the value of its parallax ($\sim$99~\%). Moreover, LP~And is a very faint star, with a magnitude of $G=17.6$~mag. As we did not apply any correction to the parallax error of such faint stars, the parallax fractional error that we used to calculate the \textit{Gaia} distance of this source could be underestimated. On the other hand, LP~And has a period of 614~days, outside the range of periods with which we derived our Mira PL relation, which could have led to this observed discrepancy. The good agreement between the derived \textit{Gaia} DR3 and PL distances for the Miras in Fig.~\ref{fig:comparison Gaia PL}, within their respective uncertainties, demonstrates the reliability of our new PL relation based on the VLBI distances for Mira variables. The distances of some of the SRs derived with the PL relation by \citet{Knapp2003} are slightly deviant from the corrected \textit{Gaia} DR3 distances, some differing by $\geq50$\,\%. There is no clear distinction between the SRas and SRbs. This discrepancy in the two methods is likely due to uncertainties related to the complex pulsation behaviour of these SRs and highlights the need for a better-constrained PL relation for these sources.
\subsection*{Comparison with the NESS catalogue}
\begin{figure}
   \includegraphics[scale=0.49]{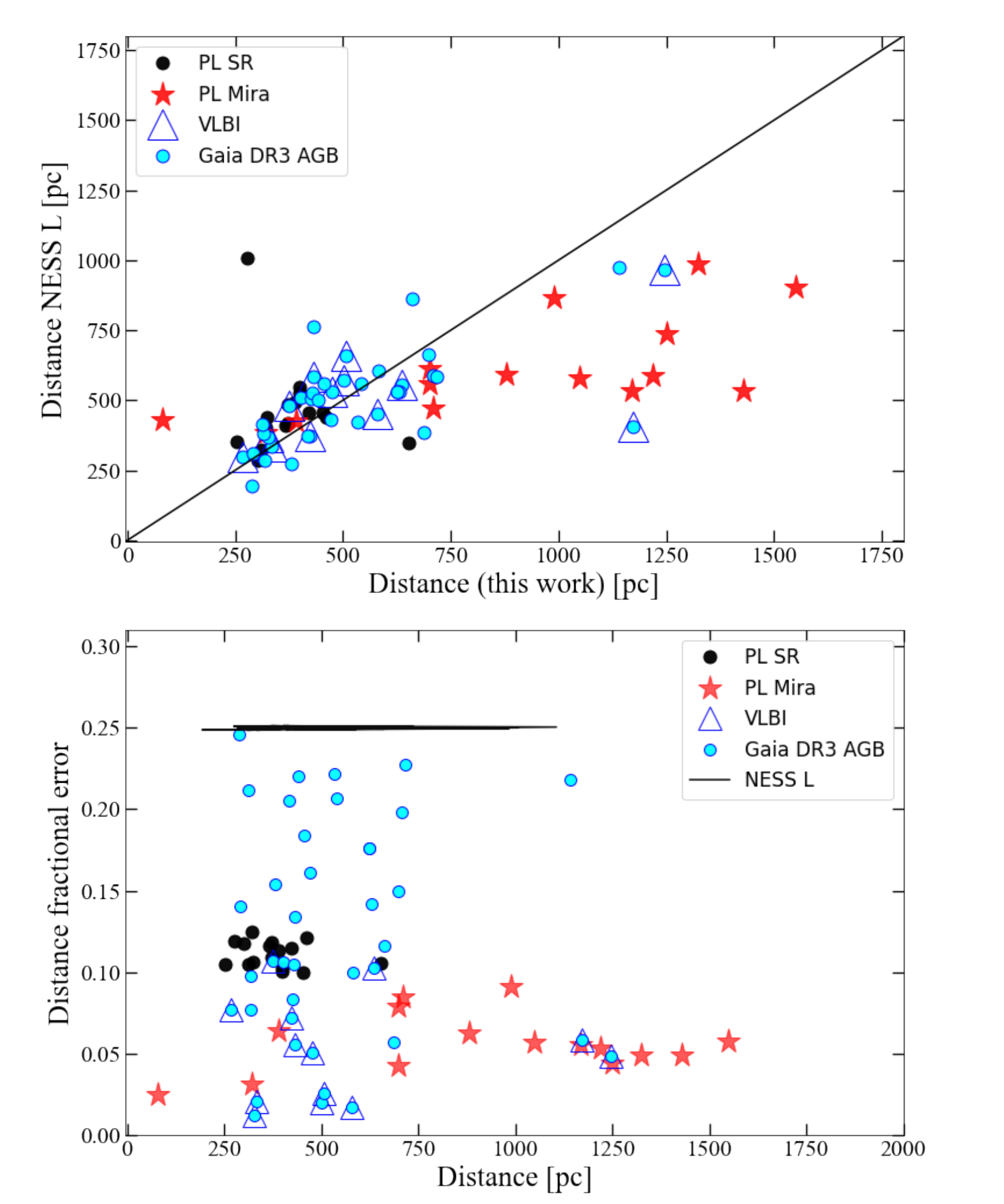}
    
    \caption{Comparison with the NESS luminosity distance. \textbf{Top}: the NESS luminosity distance (L) compared with the distances calculated with the various methods used in this work. The black line shows the 1-to-1 relation. \textbf{Bottom:} The distance fractional errors obtained in this work and those derived with the NESS luminosity distance.}
    \label{fig:NESS}
\end{figure}

A new distance catalogue for evolved stars was recently published by \citet{NESS2021} as part of the Nearby Evolved Stars Survey (NESS\footnote{\url{https://evolvedstars.space/catalogue/}}). The NESS catalogue comprises distances for more than 800 stars, including AGB stars as well as other giants and supergiants. Their distances for AGB stars  that are beyond $\sim 400\,$pc are based on a new metric derived from the luminosity probability distribution of AGB stars in the LMC. In addition, the NESS catalogue provides distances based on previous maser, \textit{Tycho–Gaia} Astrometric Solution (TGAS) of \textit{Gaia} Data Release 1 (DR1), and \textit{Hipparcos} inverted parallaxes, with parallax fractional errors within 25\,\%. About 55\,\% of the sources in the DEATHSTAR sample have distances in the NESS catalogue, mainly based on their new luminosity distance metric ($\sim 37$\,\%), \textit{Hipparcos} data ($\sim 16$\,\%), and previous maser measurements ($\sim 2$\,\%). Figure~\ref{fig:NESS} shows that the NESS luminosity distances agree reasonably well with the distances that were derived in this work, in particular for sources within $750\,$pc (in our distance). The errors on the NESS luminosity distances are fixed to 25\,\% for all sources, whereas the fractional errors on the distances that we derived change from source to source and are below 25\,\% for all the sources that are in both the NESS and the DEATHSTAR samples.
\subsection*{RUWE}
The re-normalised unit weight error or RUWE  was introduced by \citet{ruwe2018} for \textit{Gaia} DR2 and is a measure of the
goodness-of-fit of astrometric data. As previously mentioned, the RUWE and the astrometric excess noise both measure discrepancies due to photocentric motions. The astrometric excess noise is expressed as an angle with an ideal value of 0 mas for a good fit, while the RUWE is dimensionless, with an ideal value of 1.0 for well-behaved sources. A value of RUWE\,$\leq1.4$ is the criterion for good astrometric solutions. \citet{ruwe2018} obtained that 1.4 good-fit criterion by looking at the shape of the distribution of RUWE for a sample of 338\,833 sources within 100 pc of the Sun. In their work, the distribution of RUWE follows a normal distribution that peaks at approximately 1.0, but exhibits a long tail towards higher values, with a breakpoint at RUWE\,$\simeq1.4$.

\begin{figure}
   \includegraphics[scale=0.42]{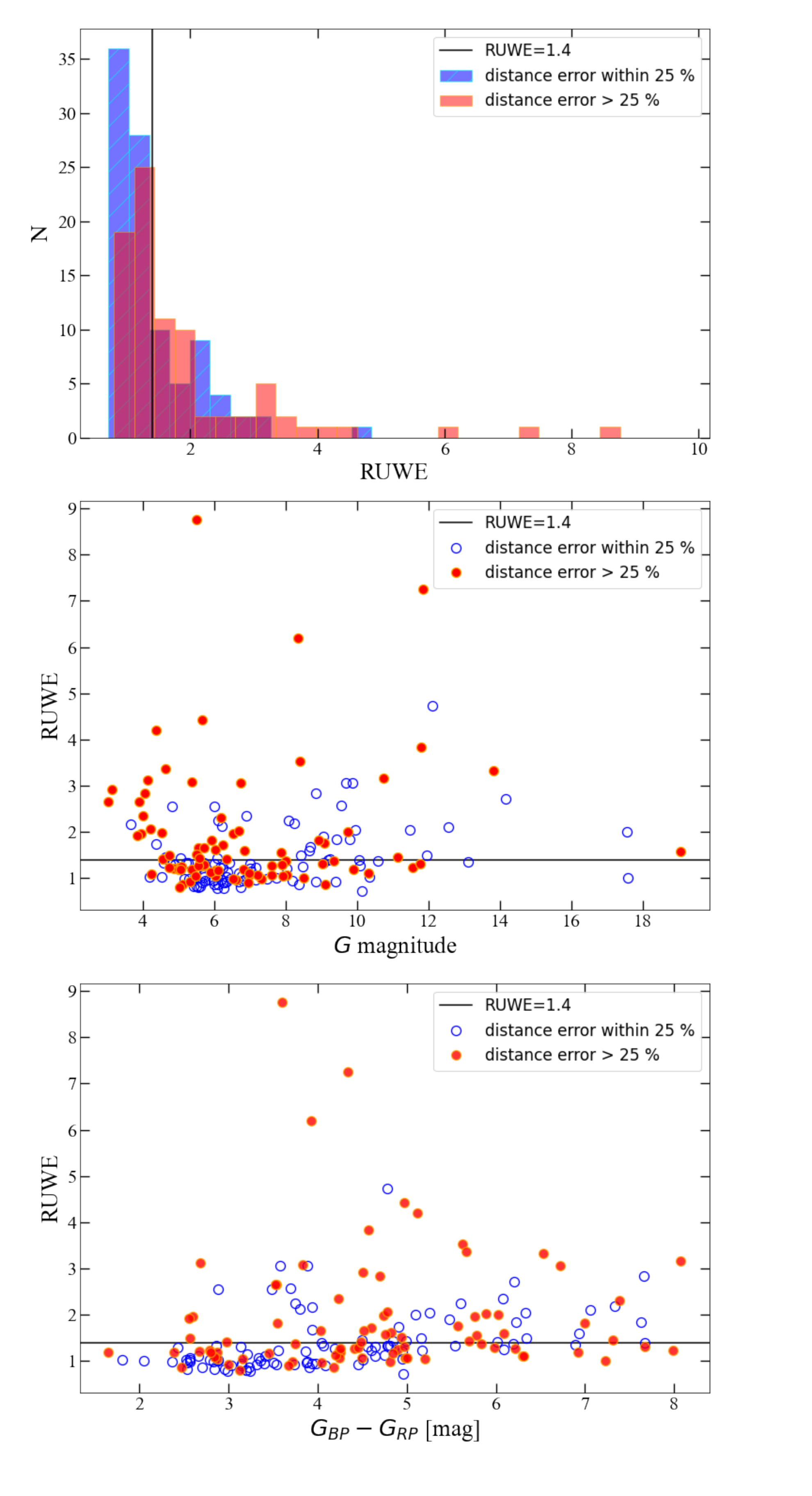}
    
    \caption{Assessment of the RUWE criterion. \textbf{Top}: Distribution of the RUWE of the sources in the DEATHSTAR sample for the 'good' (blue) and 'bad' (red) derived \textit{Gaia} DR3 distances.
    \textbf{Middle}: RUWE of the sources in the DEATHSTAR sample as a function of $G$~magnitude, where the blue open circles and the red full circles represent distances with fractional error within and above 25\,\%, respectively.
    \textbf{Bottom}: RUWE of the sources in the DEATHSTAR sample as a function of  colour. The symbols are the same as above.}
    \label{fig:ruwe}
\end{figure}

The distribution of the RUWE of the sources in the DEATHSTAR sample that have a \textit{Gaia} DR3 distance fractional error within 25\,\% has a shape similar to the distribution in \citet{ruwe2018}, as seen in Fig.~\ref{fig:ruwe}, with a median value of $\sim 1.2$ and a tail reaching a maximum RUWE of $\sim4.8$. The RUWE of about 40\,\% of these sources whose derived \textit{Gaia} distances are reliable are smaller than or equal to 1.4. Although the RUWE of the sources with distance fractional errors larger than 25\,\% can reach higher values ($\sim 8.8$), more than 20\,\% of these sources have a RUWE within the 1.4 good-fit criterion. Therefore, a RUWE value of 1.4 does not guarantee reliable distance estimates, and we caution against the use of only the RUWE to assess the quality of astrometric data from \textit{Gaia} DR3 for AGB stars. Figure~\ref{fig:ruwe} also shows that, irrespective of the RUWE, bright stars ($G<8$~mag) are dominant in the sample of sources with distance fractional error larger than 25\,\%. There is no clear trend with the distance fractional error and the colour.

\citet{ruwe2018} found that the RUWE parameter is most useful to assess the quality and reliability of the astrometric data of samples that include extremely bright, red or blue sources. However, the results of \citet{Fabricius2021} showed that the RUWE value in \textit{Gaia} DR3 for bright sources in crowded areas is strongly underestimated. The RUWE can be used to detect unresolved binaries \citep{Lindegren2021}. \citet{Stassun2021} found that, due to its high sensitivity to photocenter motions, a RUWE that is even slightly greater than 1.0 may indicate the presence of unresolved binaries. About 78\,\% of the sources in the DEATHSTAR sample have a RUWE higher than 1.0. For AGB stars, however, a high RUWE is more likely caused by saturation of the detector, the large size of the star, and/or the photocentre shift caused by convective motions on the stellar photosphere.

\section{Summary and conclusions}
\label{sec: conclusion}
A number of studies have shown that the parallaxes of AGB stars measured with \textit{Gaia} are bound to have large errors, as their intrinsic properties bring additional uncertainties to the parallax measurements \citep[circumstellar dust, colour, large size, and surface brightness variability;][]{Chiavassa2018,Xu2019,El-Badry2021}. Deriving distances from parallaxes in general is not a straightforward process, and the large errors of the parallaxes of AGB stars make it even more complicated. In this work, we assessed the \textit{Gaia} DR3 parallaxes and the corresponding distances for two samples of nearby AGB stars, the DEATHSTAR and the VLBI samples. Our main results can be summarised as follows.

The standard errors of the \textit{Gaia} DR3 parallaxes are underestimated by more than a factor of 5 for the brightest sources ($G<8$~mag), based on a comparison with the more robust VLBI parallaxes. Introducing the astrometric excess noise in the total error, as was done for \textit{Gaia} DR2, would overestimate the uncertainties. The excess noise for DR3 is higher than the DR2 excess noise for $\geq$\,60\,\% of the sources in both samples.

We inferred distances from parallaxes using a Bayesian approach that follows the procedure in \citet{BJ2015}. This method requires the use of a prior that provides information on the distances. The best prior is one that uses all available information on the sources in order to obtain the most realistic distances. We introduced the AGB prior which follows the Galactic distribution of AGB stars by \citet{Jura1990}, with a scale height of 240 pc and a scale length of 3500 pc. The most important parameter in this inference problem is the fractional error on the measured parallaxes. Our results confirmed the higher limit on the parallax fractional error of 0.2 stated by \cite{BJ2015} for good measurements. Below that limit, the posterior distribution of the distance is closely related to the likelihood of the measurements and to $1 / \varpi$. This was the case for the VLBI sample and more than half of the DEATHSTAR sample, using the corrected \textit{Gaia} DR3 parallaxes. For the remaining sources with fractional errors above 0.2, the posterior is dominated by the prior, and the errors on the distances are usually large and asymmetrical. The AGB prior, although representative of the distribution of AGB stars, is highly sensitive to the level of fractional noise, and some cases of non-convergence were observed due to the constraints imposed by the prior.
The estimated distances present large uncertainties when the fractional parallax errors are large, irrespective of the prior used. They are also less likely to be close to the values given by $1 / \varpi$. It is important to note that the measured parallax $\varpi$ is not the true parallax, and adopting the distance as the inverse of the measured parallax is not reliable.

The radiative code DUSTY was used to determine the luminosity of the VLBI sources using the better-constrained distances obtained with the VLBI parallaxes. We used the calculated luminosities to derive a new bolometric PL relation for oxygen-rich Mira variables in the Milky Way, valid for periods between 276 and 514 days. We obtained a PL relation of the form
${M_\mathrm{bol} = (-3.31\pm{0.24})\,[\mathrm{log}\, P - 2.5] + (-4.317\pm0.060)}$. The PL relation for M-type Mira variables in the Galaxy derived in this work does not significantly differ from existing bolometric PL relations for C-type Miras in the LMC and in the Milky Way \citep[e.g.][]{feast2006,White2006,White2009}.

We provided a new distance catalogue for about $200$ nearby AGB stars estimated from VLBI and corrected DR3 \textit{Gaia} parallaxes, and PL relations for Mira (this work) and SR \citep{Knapp2003} variables. Finally, we caution against the use of the RUWE  parameter as the sole measure of the quality of \textit{Gaia} DR3 astrometric data for individual AGB stars, as a RUWE below 1.4 does not guarantee reliable distance estimates.
\begin{acknowledgements}
    The authors are grateful to Kjell Eriksson for computing the MARCS models, and Sara Bladh for the helpful discussions on the stellar spectra for the SED modelling. We thank the referee for their constructive feedback on the manuscript. EDB acknowledges financial support from the Swedish National Space Agency. WV acknowledges support from the Swedish Research Council through grant No. 2020-04044. This project has received funding from the European Research Council (ERC) under the European Union’s Horizon 2020 research and innovation programme under grant agreements No. 883867 [EXWINGS]. This work presents results from the European Space Agency (ESA) space mission \textit{Gaia}. \textit{Gaia} data are being processed by the \textit{Gaia} Data Processing and Analysis Consortium (DPAC). Funding for the DPAC is provided by national institutions, in particular the institutions participating in the \textit{Gaia} MultiLateral Agreement (MLA). The \textit{Gaia} mission website is https://www.cosmos.esa.int/gaia. The \textit{Gaia} archive website is https://archives.esac.esa.int/gaia. This research made use of Astropy,\footnote{http://www.astropy.org} a community-developed core Python package for Astronomy \citep{astropy:2013, astropy:2018}. This research has made use of the VizieR catalogue access tool and the SIMBAD database, operated at CDS, Strasbourg, France.
\end{acknowledgements}
\bibliographystyle{aa} 
\bibliography{distance_AGB}
\clearpage
\begin{appendix} 
\section{\textit{Gaia} DR3 distances}
\label{Sec:appendix1}
\begin{strip}
{
\renewcommand{\arraystretch}{1.1}
 \captionof{table}{Derived \textit{Gaia} DR3 distances.}
  \begin{center}
    \begin{tabular}{lccccccc}
    \toprule
    Source & Parallax frac error$^*$& $r_\mathrm{AGB} ^\mathrm{median}\,$& $\sigma_{r\mathrm{,AGB}} ^-\,$&$\sigma_{r\mathrm{,AGB}} ^+\,$&$r_\mathrm{EDSD} ^\mathrm{median}\,$& $\sigma_{r\mathrm{,EDSD}} ^-\,$&$\sigma_{r\mathrm{,EDSD}} ^+\,$ \\
    & & [pc] & [pc] & [pc] & [pc] &[pc]&[pc] \\
     \midrule
  AA Cam & 0.13 & 489 & 60 & 79 & 480 & 75 & 57 \\
  AA Cyg & 0.27 & 1423 & 599 & 1570 & 747 & 243 & 156 \\
  AD Cyg & 0.2 & 1520 & 317 & 558 & 1202 & 243 & 179 \\
  AH Dra & 0.21 & 371 & 76 & 130 & 358 & 112 & 69 \\
  AI Vol & 0.1 & 583 & 56 & 70 & 573 & 67 & 53 \\
  AM Cen & 0.12 & 897 & 103 & 134 & 854 & 113 & 91 \\
  AP Lyn & 0.15 & 495 & 72 & 101 & 480 & 90 & 66 \\
  AQ And & 0.1 & 768 & 72 & 89 & 756 & 84 & 69 \\
  BD+06 319 & 0.18 & 312 & 52 & 80 & 310 & 76 & 52 \\
  BK Vir & 0.17 & 242 & 39 & 59 & 242 & 58 & 39 \\
  BL Ori & 0.33 & 1931 & 970 & 1570 & 751 & 291 & 181 \\
  BM Gem & 0.18 & 1243 & 201 & 293 & 1129 & 213 & 158 \\
  BW Cam & 0.26 & 1104 & 274 & 505 & 893 & 251 & 169 \\
  BX Cam & 0.14 & 554 & 75 & 104 & 537 & 92 & 69 \\
  BX Eri & 0.14 & 439 & 58 & 79 & 433 & 75 & 56 \\
  CL Mon & 0.27 & 1563 & 471 & 989 & 1042 & 281 & 194 \\
  CS Dra & 0.16 & 457 & 69 & 99 & 450 & 92 & 66 \\
  CSS2 41 & 0.53 & 3792 & 700 & 497 & 2398 & 501 & 388 \\
  CW Cnc & 0.11 & 262 & 27 & 35 & 261 & 34 & 27 \\
  CZ Hya & 0.2 & 1514 & 253 & 364 & 1366 & 266 & 198 \\
  DK Vul & 0.27 & 1625 & 596 & 1345 & 926 & 269 & 181 \\
  DR Ser & 0.18 & 1200 & 223 & 369 & 1010 & 194 & 144 \\
  DY Gem & 0.25 & 1217 & 357 & 838 & 862 & 241 & 163 \\
  EP Aqr & 0.34 & 500 & 299 & 700 & 305 & 369 & 145 \\
  EP Vul & 0.3 & 2124 & 996 & 1522 & 871 & 281 & 185 \\
  FU Mon & 0.13 & 795 & 99 & 132 & 755 & 112 & 86 \\
  FV Boo & 0.21 & 885 & 155 & 227 & 857 & 204 & 144 \\
  GI Lup & 0.17 & 1019 & 169 & 254 & 917 & 175 & 129 \\
  GX Mon & 0.23 & 727 & 203 & 522 & 570 & 179 & 114 \\
  GY Aql & 0.59 & 1965 & 952 & 1322 & 731 & 396 & 239 \\
  HS UMa & 0.17 & 332 & 52 & 74 & 332 & 75 & 52 \\
  HU Pup & 0.44 & 3390 & 895 & 758 & 1806 & 435 & 324 \\
  HV Cas & 0.14 & 1205 & 163 & 225 & 1112 & 171 & 133 \\
  IK Tau & 0.19 & 289 & 54 & 88 & 281 & 79 & 50 \\
  IRAS 15194-5115 & 0.14 & 696 & 93 & 129 & 656 & 104 & 79 \\
  IRC -10401 & 0.64 & 3383 & 1166 & 798 & 1121 & 414 & 280 \\
  IRC+10365 & 0.16 & 519 & 82 & 122 & 488 & 96 & 70 \\
  IRC+60041 & 0.14 & 1300 & 175 & 241 & 1179 & 172 & 136 \\
  IRC$-$30398 & 0.54 & 2172 & 1018 & 1340 & 812 & 383 & 237 \\
  L2 Pup & 0.25 & 1043 & 942 & 1579 & 98 & 190 & 37 \\
  LP And & 0.1 & 428 & 40 & 50 & 423 & 48 & 39 \\
  NP Pup & 0.13 & 586 & 70 & 91 & 570 & 83 & 64 \\
  NSV 17351 & 1.66 & 3873 & 687 & 447 & 2402 & 547 & 424 \\
  NSV 24833 & 0.24 & 1155 & 277 & 513 & 928 & 244 & 167 \\
  OH 56.1 +2.1 & 0.65 & 3953 & 616 & 392 & 2601 & 528 & 414 \\
    \bottomrule
    \multicolumn{4}{l}{\footnotesize $^*$ corrected parallax fractional error } \\
     \multicolumn{4}{l}{\footnotesize $r_\mathrm{prior} ^\mathrm{median}$ the median distance derived with the AGB or EDSD prior. } \\
    \multicolumn{4}{l}{\footnotesize $\sigma_{r\mathrm{,prior}} ^{-/+}$ the lower/upper uncertainties of the distance with the AGB or EDSD prior.} \\
    \end{tabular}
    \label{table: dist Append1}
  \end{center}
}
\end{strip}

{
\renewcommand{\arraystretch}{1.18}
\begin{table*}
  \renewcommand\thetable{A.1}
   
\caption{continued.}
  \begin{center}
    \begin{tabular}{lccccccc}
   \toprule
    Source & Parallax frac error& $r_\mathrm{AGB} ^\mathrm{median}\,$& $\sigma_{r\mathrm{,AGB}} ^-\,$&$\sigma_{r\mathrm{,AGB}} ^+\,$&$r_\mathrm{EDSD} ^\mathrm{median}\,$& $\sigma_{r\mathrm{,EDSD}} ^-\,$&$\sigma_{r\mathrm{,EDSD}} ^+\,$ \\
    & & [pc] & [pc] & [pc] & [pc] &[pc]&[pc] \\
     \midrule
  OZ Gem & 0.71 & 1913 & 568 & 866 & 1339 & 437 & 307 \\
 PQ Cep & 0.08 & 631 & 45 & 54 & 625 & 51 & 45 \\
  QX Pup & 5.37 & 3820 & 706 & 481 & 2392 & 559 & 433 \\
  R And & 0.32 & 655 & 253 & 627 & 478 & 258 & 137 \\
  R Aql & 0.2 & 308 & 77 & 268 & 266 & 85 & 52 \\
  R Aqr & 0.68 & 631 & 263 & 448 & 618 & 434 & 255 \\
  R Cas & 0.19 & 207 & 41 & 74 & 198 & 60 & 37 \\
  R Cnc & 0.24 & 347 & 97 & 235 & 311 & 141 & 75 \\
  R Crt & 0.17 & 237 & 40 & 64 & 234 & 59 & 39 \\
  R Cyg & 0.09 & 555 & 45 & 54 & 548 & 52 & 43 \\
  R For & 0.05 & 507 & 24 & 27 & 507 & 27 & 24 \\
  R Gem & 0.19 & 847 & 159 & 256 & 763 & 174 & 123 \\
  R Hor & 0.19 & 260 & 48 & 77 & 258 & 75 & 47 \\
  R Hya & 0.37 & 595 & 350 & 730 & 355 & 386 & 169 \\
  R LMi & 0.22 & 339 & 73 & 127 & 333 & 118 & 70 \\
  R Lep & 0.14 & 471 & 64 & 88 & 462 & 82 & 60 \\
  R Lyn & 0.13 & 880 & 105 & 138 & 852 & 123 & 97 \\
  R Peg & 0.24 & 450 & 104 & 186 & 430 & 156 & 93 \\
  R Scl & 0.16 & 408 & 61 & 86 & 409 & 87 & 61 \\
  R UMa & 0.26 & 638 & 145 & 245 & 608 & 207 & 130 \\
  R Vol & 0.11 & 662 & 68 & 86 & 649 & 81 & 65 \\
  RR Aql & 0.15 & 513 & 73 & 102 & 492 & 88 & 65 \\
  RS And & 0.25 & 562 & 175 & 491 & 443 & 176 & 102 \\
  RS CrA & 0.07 & 1467 & 97 & 112 & 1437 & 105 & 91 \\
  RT Cap & 0.13 & 564 & 69 & 92 & 551 & 84 & 65 \\
  RT Sco & 0.31 & 2592 & 1453 & 1370 & 724 & 271 & 169 \\
  RT Vir & 0.29 & 343 & 107 & 233 & 338 & 220 & 104 \\
  RV Aqr & 0.09 & 586 & 50 & 61 & 580 & 59 & 49 \\
  RV Cam & 0.22 & 490 & 129 & 324 & 413 & 139 & 85 \\
  RV Cyg & 0.3 & 1124 & 426 & 1051 & 698 & 261 & 162 \\
  RW LMi & 0.08 & 319 & 22 & 27 & 319 & 27 & 23 \\
  RW Lep & 0.16 & 419 & 63 & 91 & 407 & 81 & 58 \\
  RX Boo & 0.19 & 181 & 35 & 59 & 180 & 59 & 35 \\
  RX Lac & 0.28 & 596 & 228 & 724 & 421 & 204 & 110 \\
  RY Dra & 0.1 & 401 & 38 & 47 & 400 & 46 & 38 \\
  RY Mon & 0.11 & 875 & 96 & 124 & 835 & 106 & 85 \\
  RZ Peg & 0.11 & 1275 & 128 & 161 & 1225 & 141 & 115 \\
  RZ Sgr & 0.12 & 432 & 50 & 66 & 426 & 62 & 49 \\
  S Aur & 0.2 & 1244 & 267 & 494 & 1000 & 217 & 156 \\
  S Cas & 0.2 & 965 & 195 & 331 & 833 & 194 & 136 \\
  S Cep & 0.18 & 534 & 93 & 144 & 504 & 117 & 80 \\
  S CrB & 0.23 & 443 & 98 & 167 & 433 & 153 & 93 \\
  S Crt & 0.25 & 548 & 123 & 211 & 530 & 187 & 114 \\
  S Dra & 0.36 & 707 & 231 & 448 & 603 & 298 & 170 \\
  S Lyr & 1.34 & 3473 & 899 & 706 & 1770 & 496 & 367 \\
  S Pav & 0.57 & 1154 & 626 & 971 & 533 & 458 & 267 \\
  S Sct & 0.12 & 438 & 50 & 65 & 425 & 59 & 46 \\
  S Ser & 0.38 & 1079 & 262 & 414 & 998 & 339 & 225 \\
    \hline
    \end{tabular}
    \label{table: dist Append2}
  \end{center}
\end{table*}
}
{
\renewcommand{\arraystretch}{1.18}
\begin{table*}
  \renewcommand\thetable{A.1}

\caption{continued.}
  \begin{center}
    \begin{tabular}{lccccccc}
   \toprule
    Source & Parallax frac error& $r_\mathrm{AGB} ^\mathrm{median}\,$& $\sigma_{r\mathrm{,AGB}} ^-\,$&$\sigma_{r\mathrm{,AGB}} ^+\,$&$r_\mathrm{EDSD} ^\mathrm{median}\,$& $\sigma_{r\mathrm{,EDSD}} ^-\,$&$\sigma_{r\mathrm{,EDSD}} ^+\,$ \\
    & & [pc] & [pc] & [pc] & [pc] &[pc]&[pc] \\
     \midrule
  SS Vir & 0.22 & 583 & 112 & 173 & 576 & 167 & 108 \\
  ST Cam & 0.15 & 625 & 91 & 129 & 597 & 109 & 81 \\
  ST Her & 0.18 & 324 & 57 & 86 & 319 & 82 & 54 \\
  ST Sco & 0.22 & 786 & 197 & 438 & 637 & 178 & 118 \\
  ST Sgr & 0.19 & 608 & 127 & 237 & 534 & 136 & 91 \\
  SU Vel & 0.16 & 417 & 68 & 103 & 396 & 85 & 59 \\
  SV Aqr & 0.16 & 445 & 65 & 90 & 443 & 89 & 64 \\
  SV Peg & 0.35 & 1000 & 504 & 1179 & 501 & 286 & 152 \\
  SW Vir & 0.43 & 491 & 245 & 487 & 433 & 411 & 205 \\
  SY Aql & 0.2 & 973 & 203 & 358 & 824 & 191 & 135 \\
  SY Scl & 0.48 & 1157 & 261 & 382 & 1164 & 387 & 264 \\
  SZ Car & 0.07 & 689 & 45 & 52 & 681 & 50 & 44 \\
  SZ Dra & 0.13 & 470 & 57 & 75 & 460 & 70 & 54 \\
  T Ari & 0.42 & 689 & 270 & 533 & 554 & 342 & 185 \\
  T Cam & 0.21 & 623 & 136 & 248 & 554 & 155 & 102 \\
  T Cep & 0.34 & 1315 & 889 & 1451 & 338 & 322 & 136 \\
  T Cet & 0.44 & 465 & 193 & 377 & 473 & 388 & 199 \\
  T Dra & 0.18 & 901 & 146 & 214 & 849 & 175 & 126 \\
  T Ind & 0.23 & 671 & 139 & 227 & 634 & 188 & 122 \\
  T Lep & 0.18 & 356 & 61 & 94 & 347 & 85 & 57 \\
  T Lyr & 0.08 & 427 & 33 & 38 & 423 & 38 & 31 \\
  T Mic & 0.25 & 272 & 84 & 240 & 241 & 131 & 63 \\
  T Sgr & 0.96 & 2839 & 1015 & 1051 & 1251 & 448 & 310 \\
  TT Cen & 0.13 & 1182 & 152 & 208 & 1083 & 154 & 121 \\
  TT Cyg & 0.07 & 671 & 43 & 49 & 664 & 47 & 42 \\
  TT Tau & 0.09 & 671 & 59 & 73 & 659 & 68 & 56 \\
  TU Gem & 0.38 & 2129 & 884 & 1350 & 989 & 340 & 224 \\
  TV Dra & 0.18 & 541 & 90 & 134 & 523 & 119 & 83 \\
  TW Hor & 0.17 & 481 & 75 & 109 & 475 & 104 & 73 \\
  TW Oph & 0.25 & 966 & 356 & 1276 & 613 & 205 & 128 \\
  TW Peg & 0.17 & 278 & 48 & 77 & 269 & 68 & 44 \\
  TX Cam & 0.13 & 292 & 35 & 47 & 287 & 44 & 33 \\
  TX Psc & 0.28 & 339 & 101 & 223 & 327 & 193 & 94 \\
  TY Dra & 0.14 & 699 & 89 & 120 & 681 & 110 & 83 \\
  TZ Aql & 0.13 & 524 & 64 & 84 & 511 & 78 & 60 \\
  U Ant & 0.14 & 294 & 40 & 54 & 288 & 50 & 38 \\
  U Cam & 0.13 & 630 & 77 & 102 & 606 & 89 & 70 \\
  U Cyg & 0.06 & 687 & 37 & 41 & 681 & 41 & 35 \\
  U Her & 0.17 & 453 & 73 & 110 & 442 & 99 & 69 \\
  U Lyn & 0.19 & 925 & 161 & 243 & 855 & 185 & 133 \\
  U Men & 0.09 & 317 & 28 & 34 & 315 & 33 & 27 \\
  UU Aur & 0.43 & 1295 & 642 & 1212 & 609 & 344 & 194 \\
  UX And & 0.35 & 1168 & 471 & 1027 & 700 & 301 & 181 \\
  UX Cyg & 0.29 & 1682 & 527 & 1031 & 1081 & 304 & 209 \\
  UX Dra & 0.23 & 567 & 135 & 254 & 510 & 168 & 105 \\
  UY Cen & 0.18 & 718 & 127 & 200 & 660 & 145 & 103 \\
  UY Cet & 0.22 & 455 & 90 & 143 & 454 & 142 & 89 \\
  V Aql & 0.26 & 1045 & 535 & 2005 & 488 & 193 & 113 \\
    \hline
    \end{tabular}
    \label{table: dist Append3}
  \end{center}
\end{table*}
}
{
\renewcommand{\arraystretch}{1.18}
\begin{table*}
  \renewcommand\thetable{A.1}
   
\caption{continued.}
  \begin{center}
    \begin{tabular}{lccccccc}
   \toprule
    Source & Parallax frac error& $r_\mathrm{AGB} ^\mathrm{median}\,$& $\sigma_{r\mathrm{,AGB}} ^-\,$&$\sigma_{r\mathrm{,AGB}} ^+\,$&$r_\mathrm{EDSD} ^\mathrm{median}\,$& $\sigma_{r\mathrm{,EDSD}} ^-\,$&$\sigma_{r\mathrm{,EDSD}} ^+\,$ \\
    & & [pc] & [pc] & [pc] & [pc] &[pc]&[pc] \\
     \midrule
  V CrB & 0.11 & 836 & 81 & 99 & 829 & 97 & 79 \\
  V Cyg & 0.2 & 623 & 136 & 266 & 538 & 141 & 94 \\
  V Hya & 0.25 & 529 & 131 & 248 & 484 & 177 & 107 \\
  V Tel & 0.22 & 501 & 117 & 224 & 451 & 145 & 91 \\
  V1302 Cen & 0.11 & 892 & 91 & 114 & 856 & 100 & 81 \\
  V1426 Cyg & 0.16 & 709 & 113 & 168 & 657 & 128 & 92 \\
  V1942 Sgr & 0.13 & 650 & 84 & 115 & 622 & 97 & 75 \\
  V1943 Sgr & 0.38 & 929 & 561 & 1062 & 394 & 366 & 170 \\
  V1968 Cyg & 0.46 & 2669 & 1042 & 1163 & 1135 & 380 & 258 \\
  V365 Cas & 0.2 & 1010 & 206 & 355 & 858 & 194 & 137 \\
  V384 Per & 2.64 & 2712 & 1087 & 1123 & 991 & 463 & 310 \\
  V386 Cep & 0.22 & 2561 & 558 & 849 & 1820 & 337 & 257 \\
  V460 Cyg & 0.37 & 1376 & 649 & 1300 & 661 & 310 & 182 \\
  V466 Per & 0.13 & 685 & 84 & 112 & 655 & 96 & 75 \\
  V637 Per & 0.26 & 1227 & 307 & 568 & 969 & 265 & 181 \\
  V644 Sco & 0.21 & 1608 & 382 & 779 & 1179 & 246 & 180 \\
  V688 Mon & 0.29 & 1420 & 622 & 1558 & 1687 & 1100 & 720 \\
  V821 Her & 0.36 & 1846 & 916 & 1487 & 731 & 309 & 189 \\
  V837 Her & 0.8 & 3225 & 988 & 858 & 1499 & 456 & 326 \\
  V996 Cen & 0.11 & 578 & 59 & 74 & 563 & 67 & 55 \\
  VX And & 0.08 & 619 & 49 & 59 & 612 & 56 & 48 \\
  VX Aql & 0.25 & 2394 & 673 & 1092 & 1479 & 318 & 234 \\
  VY UMa & 0.17 & 444 & 71 & 104 & 437 & 99 & 68 \\
  W Aql & 0.13 & 380 & 49 & 68 & 369 & 60 & 46 \\
  W CMa & 0.18 & 777 & 148 & 246 & 681 & 150 & 106 \\
  W Leo & 0.28 & 961 & 195 & 294 & 943 & 278 & 187 \\
  W Ori & 0.41 & 926 & 391 & 800 & 602 & 333 & 187 \\
  W Pic & 0.08 & 594 & 43 & 49 & 590 & 48 & 42 \\
  WY Cas & 0.17 & 1141 & 196 & 302 & 1001 & 190 & 141 \\
  WZ Cas & 0.07 & 465 & 33 & 38 & 460 & 37 & 32 \\
  X Cnc & 0.36 & 754 & 238 & 455 & 641 & 298 & 174 \\
  X Her & 0.17 & 137 & 23 & 34 & 136 & 34 & 22 \\
  X Hya & 0.23 & 482 & 115 & 221 & 441 & 153 & 93 \\
  X TrA & 0.28 & 941 & 505 & 1683 & 430 & 210 & 114 \\
  X Vel & 0.11 & 631 & 65 & 82 & 615 & 75 & 61 \\
  XZ Vel & 0.15 & 1151 & 162 & 228 & 1049 & 165 & 128 \\
  Y CVn & 0.29 & 397 & 110 & 213 & 397 & 214 & 110 \\
  Y Hya & 0.23 & 551 & 130 & 248 & 495 & 163 & 101 \\
  Y Lib & 0.22 & 1055 & 196 & 294 & 985 & 237 & 167 \\
  Y Lyn & 0.21 & 422 & 92 & 168 & 396 & 128 & 79 \\
  Y Pav & 0.41 & 980 & 307 & 548 & 797 & 336 & 209 \\
  Y Scl & 0.3 & 406 & 120 & 244 & 398 & 229 & 115 \\
  Y Tau & 0.2 & 782 & 169 & 318 & 669 & 168 & 115 \\
  Y Tel & 0.18 & 442 & 77 & 118 & 426 & 103 & 70 \\
  Y UMa & 0.17 & 348 & 57 & 83 & 346 & 83 & 56 \\
  Z Ant & 0.28 & 1106 & 291 & 540 & 883 & 269 & 178 \\
  Z Psc & 0.18 & 660 & 107 & 157 & 638 & 140 & 99 \\
     $\theta$ Aps & 0.23 & 203 & 75 & 886 & 157 & 88 & 40 \\
  $\chi$ Cyg & 0.24 & 1272 & 1058 & 2327 & 209 & 115 & 55 \\
    \hline
    \end{tabular}
    \label{table: dist Append4}
  \end{center}
\end{table*}
}
\FloatBarrier
\section{SED fitting with DUSTY}
\label{sec: Append dusty}
The spectral energy distributions (SEDs) of the sources in the VLBI and DEATHSTAR samples were modelled using the radiative transfer code DUSTY \citep{Ivezic1999}. A central star is surrounded by a dusty shell following a $r^{-2}$-density law, and with an outer to inner radius fraction of 10$^4$. All dust grains were assumed to have a radius $a$\,=\,0.1\,$\mu$m. The modelled VLBI sources are M-type stars for which the grains were assumed to be silicate-type with the optical properties from \citet{Justtanont1992}. We used the high-resolution MARCS\footnote{\url//marcs.astro.uu.se/} model atmospheres \citep[][private communication]{Marcs2008} as input  stellar spectra for the M-type stars. The corresponding stellar temperature ranges from 2400 to 3600~K. The chosen models have a surface gravity, log\,$g$, of $-0.5$, which is within the range of the values of $\mathrm{log\,}g$ obtained from the 3D hydrodynamical models of AGB stars of \citet{Freytag2017} and \citet{ahmad2022}; and a microturbulence of 0.5 km s$^{-1}$ as in \citet{Olofsson2002}. The resolution of the MARCS spectra was reduced by convolving them with a Gaussian kernel with a standard deviation of 250. The diluted spectra were then re-gridded. As the MARCS spectra end at $20\,\mu$m, they were extrapolated to longer wavelengths, up to 3.6\,cm, with a Rayleigh-Jeans tail.
\\For the carbon stars, the dust grains were assumed to be amorphous carbons with optical properties derived by \citet{Rouleau1991}. We used the COMARCS\footnote{\url{http://stev.oapd.inaf.it/atm/lrspe.html}} model atmospheres \citep{Aringer2019} for the stellar spectra of the carbon stars, with temperatures between 2500 and 3300~K, and the same $\mathrm{log}\,g$ and microturbulence as the MARCS models. The type of dust and input stellar spectra of the S-type stars were either similar to the C- or the M-type stars, depending on their spectral class or colour, as in \citet{Ramstedt2006}.
A large grid of radiative transfer models with varying temperature, with steps of $100\,$K ($T_\star$ =2400~--~3600 K and 2500~--~3300 K for the oxygen- and carbon-rich models, respectively); dust temperature at the inner radius of the dust shell, $T_{\rm{d}}$, ranging from 600 to 1200 K for the oxygen-rich models, and up to 1300 K for the carbon-rich models, with steps of $100\,$K; and dust optical depth at 10\,$\mu$m ($\tau_{10}=0.01-5.00$, with steps of 0.01) was constructed. We made use of the scaling properties of the dust radiative transfer in a spherically symmetric envelope to determine either the luminosity, knowing the distance, or the other way around. We used the former to derive a new PL relation with the VLBI sources using their distances obtained from maser parallaxes (Sect.~\ref{sec: PL}). The latter was used to determine the distances of the sources whose \textit{Gaia} DR3 distances are not reliable or for direct comparison with the \textit{Gaia} distances (Sect.~\ref{sec: cat}).\\

For each star, the DUSTY input parameters, which are $T_{\star}$, $T_{\rm{d}}$, and $\tau_{10}$, and the bolometric stellar luminosity, $L_{\star}$, or the distance, $r$, were constrained by photometric flux densities collected from various online catalogues. The data consist of $G$, $G_{BP}$, and $G_{RP}$ fluxes from \textit{Gaia} DR3; $J$-, $H$-, and $K_{s}$-band fluxes from 2MASS (quality flag \textit{rd\_flg} between 1 and 3); 3.5 and 4.9 $\mu$m fluxes ($L$ and $M$ bands, respectively) from DIRBE; 12, 25, 60, 100 $\mu$m fluxes from the \textit{IRAS} point source catalogue (quality flag 3), and fluxes at $8.6$ up to $160\,\mu$m from Akari (quality flag 3) collected from the ViZier photometry viewer\footnote{\url{http://vizier.unistra.fr/vizier/sed/}} online tool. We did not include WISE data as AGB stars are known to saturate the WISE photometric instruments, in particular in the \textit{W1} and \textit{W2} bands, requiring additional calibrations to correct for residual biases \citep{Lian2014}. The fluxes were corrected for interstellar extinction using the visual extinction coefficient, $A_V$, given by Eq.(8) in \citet{Martin1992}. The  $A_\lambda/A_V$ ratios were taken from \citet{Schlegel1998} and interpolated for the relevant wavelengths. Extinction beyond 3.5\,$\mu$m was neglected. \\

\citet{Ramstedt2008} assessed the effect of variability on the flux using several measurements from different epochs. Their results show that the variability of the star, which is particularly important at short wavelengths, can be accounted for in the uncertainty assumed for each flux point. Uncertainties of 60\,\%, 50\,\%, and 40\,\% were assigned to fluxes in the $J$, $H$, and $K$ bands, respectively. For the $L$ and $M$ bands, a flux uncertainty of 30\,\% was assumed. The properties of variability at longer wavelengths are less clear due to scarce data. A 20 \% calibration uncertainty was assigned to flux points beyond 5\,$\mu$m. The \textit{Gaia} bands were not covered in \citet{Ramstedt2008}. As the amplitude of the variation is the largest at short wavelengths, we assumed an uncertainty of 75\% for the fluxes measured at the three \textit{Gaia} bands.

\begin{figure*}[hbt!]
  \includegraphics[page=1,width=1.00\textwidth]{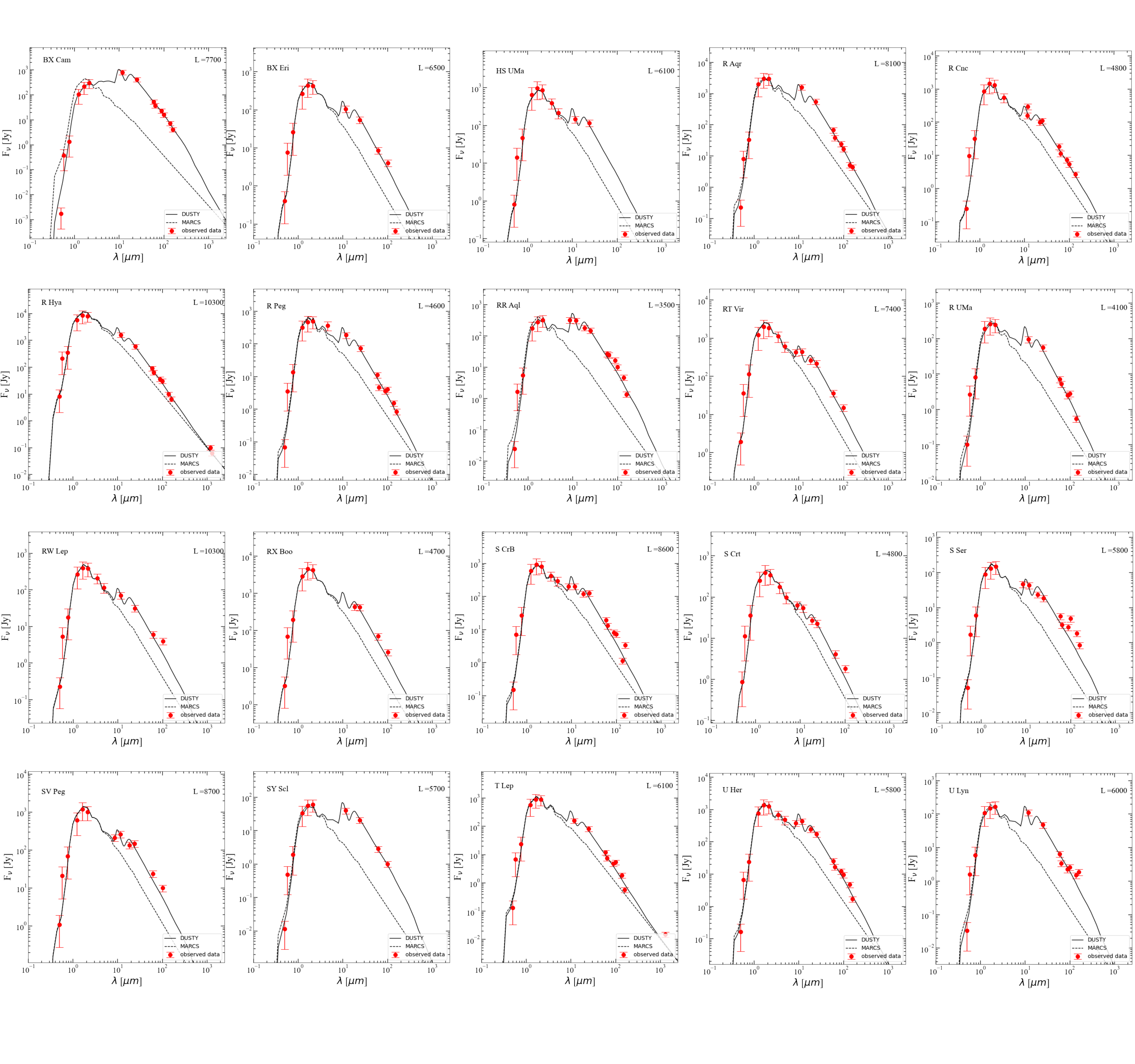}
   
    \caption{SED fitting of the stellar and dust emission with DUSTY for the sources in the VLBI sample. The name of each source is given in the upper left corner of each panel, and the derived luminosity is given in the upper right corner, in solar luminosity.}
    \label{fig:SED1}
\end{figure*}

The model grid SEDs were scaled according to the median distance, $r_\mathrm{median}$, derived using the VLBI parallax and the AGB prior (or the luminosity derived with our new PL relation)  until a satisfactory fit to the observational data was obtained. The best-fit to the DUSTY input parameters (see Fig.~\ref{fig:SED1}) was found using the same $\chi^2$-minimization strategy as in \citet{Ramstedt2009}, for example. Thereby, $T_{\star}$, $T_{\rm{d}}$, $\tau_{10}$, and $L_{\star}$ (or $r$) were estimated for each source.
As the derived VLBI distance (or the luminosity calculated using the PL relation derived in Sect.~\ref{sec: PL}) follows a distribution centred on $r_\mathrm{median}$ ($L_{\star} ^\mathrm{median}$) with known lower and upper limits, we checked whether the best fit model significantly changed for the extreme distances (luminosities). We then optimised the flux scaling with the now-known $T_{\star}$, $T_{\rm{d}}$, and $\tau_{10}$ parameters. We created a posterior distribution on the luminosity $L_{\star}$ (distance $r$) by applying the weighted bootstrap statistical method which consists of randomly selecting a distance (luminosity) from the weighted distance (luminosity) distribution and fitting it to a sub-sample of the observed photometric data selected at random. This process was repeated $4\,000$ times, which allowed us to derive a median luminosity, $L_{\star} ^\mathrm{median}$ (distance, $r_{\mathrm{median}}$), along with the 16 and 84 percentiles of the generated luminosity (distance) distribution as its confidence interval. The derived parameters are listed in Tables~\ref{table: RT} and \ref{table: RT Appendix1}.

The SEDs of a number of the sources that we modelled were poorly fitted. This is particularly obvious for HU~Pup, QX~Pup, and UX~Cyg in the VLBI sample. The poor fit of QX~Pup could be explained by its complex transition nature. As previously mentioned, QX~Pup is classified as an OH/IR star. It is the central star of the bipolar nebula OH231.8, and is part of a system with a binary companion and a rotating circumbinary disk \citep{QXPup2022}. The poor fit of UX~Cyg could be due to its unusually turbulent envelope \citep{etoka2018}. There is no record of signs of irregularities for the SRa variable HU~Pup in the literature. For sources such as S~Ser and NSV~17351, the data points were not properly fitted in the long wavelength part of the SED. In fact, the SED of a few sources exhibit a second bump at longer wavelengths (e.g. NSV~17351, GI~Lup, RT~Sco, TT~Cen), which is usually observed in sources with a detached shell. For such sources, the SED is usually fitted with two black bodies, which was not attempted in this paper.
\subsubsection*{Comparison between black body and model atmosphere}
We compared the results of the SED fitting when using a black body as stellar radiation input with the results obtained with model atmospheres. We found that the results between the two methods differ the most for oxygen-rich stars due to the strong TiO absorption in their spectra at short wavelengths. Overall, the luminosity is lower by about 25\,\%  with the MARCS models compared to with a black body. When using model atmospheres, the dust condensation temperature is systematically lower than what one obtains with a black body, by up to 35\,\%, while the corresponding effective temperatures tend to be higher, by up to 30\,\%, for oxygen-rich stars. The optical depths at 10~$\mu$m obtained with the two methods are in good agreement in the low-$\tau_{10}$ regime ($\tau_{10} < 0.8$). Above that, the optical depths derived using model atmospheres tend to be smaller.
{
\renewcommand{\arraystretch}{1.2}
\begin{table*}
 
\caption{DUSTY results for the Miras in the DEATHSTAR sample for which distances were derived using our new PL relation. }
  \begin{center}
    \begin{tabular}{lccccccccccc}
    \toprule
    Source & $P$ & dust$^*$ & $T_\star$ &$T_\mathrm{dust}$& $\tau_{10}$ & $L_\star ^\mathrm{median}\,$& $\sigma_{L_\star} ^\mathrm{-}\,$& $\sigma_{L_\star} ^\mathrm{+}\,$ &$r_\mathrm{PL(M)}$ & $\sigma _{r,\mathrm{PL(M)}}^-$ &$\sigma _{r,\mathrm{PL(M)}}^+$ \\
     & [days] & & [K]& [K] & & [L$_\odot$] & [L$_\odot$] & [L$_\odot$] &[pc]&[pc]&[pc] \\
     \midrule
\textit{C-type}\\
AFGL 3068 & 696& AMC & 2500& 800& 1.2 & 11890& 1013& 1108 & 1220 & 60 & 70\\
AI Vol & 511& AMC & 2500& 1300& 0.28 & 7897& 525& 563 & 730 & 60 & 50\\
CL Mon & 493& AMC & 3100& 1300& 0.05 & 7532& 485& 518& 910 & 30 & 40\\
CZ Hya & 442& AMC & 2500& 1200& 0.09 & 6518& 376& 399& 1470 & 60 & 70 \\
HV Cas & 527& AMC & 3100& 1300& 0.09 & 8226& 563& 604 & 1320 & 70 & 80\\
IRAS 15194-5115 & 576& AMC & 3000& 1200& 0.36 & 9254& 683& 738& 670 & 40 & 40\\
IRC+10216 & 630& AMC & 3100& 1300& 0.5 & 10420& 826& 897 & 190 & 20 & 20\\
IRC+60041 & 280& AMC & 2500& 1300& 0.07 & 3561& 104& 107& 640 & 50 & 60 \\
LP And & 614& AMC & 2500& 1300& 3& 10071& 783& 849& 1760 & 120 & 100\\
PQ Cep & 442& AMC & 2500& 1300& 0.04 & 6517& 376& 399 & 1080 & 80 & 130\\
R For & 386& AMC & 2500& 1300& 0.06 & 5447& 268& 282 & 650 & 50 & 50\\
R Lep & 445& AMC & 3300& 1100& 0.02 & 6576& 382& 405 & 380 & 20 & 10\\
R Vol & 453& AMC & 2500& 1300& 0.13 & 5076& 233& 244 & 840 & 60 & 50\\
RV Aqr & 453& AMC & 3100& 1300& 0.1 & 6733& 398& 423 & 600 & 30 & 30\\
RZ Peg & 437& AMC & 2700& 900& 0.01 & 6420& 365& 388 & 1380 & 70 & 80\\
S Cep & 484& AMC & 3100& 1300& 0.03 & 7350& 465& 496 & 390 & 20 & 20\\
T Dra & 422& AMC & 2500& 1300& 0.07 & 6134& 337& 356& 670 & 50 & 30\\
U Cyg & 463& AMC & 3000& 800& 0.02 & 6930& 419& 446 & 870 & 50 & 60\\
V CrB & 358& AMC & 3100& 1300& 0.02 & 4930& 220& 230 & 640 & 20 & 30\\
V Cyg & 421& AMC & 2500& 1200& 0.07 & 6116& 335& 354& 420 & 10 & 20\\
V358 Lup & 632& AMC & 2500& 1300& 0.75 & 10464& 831& 903& 1050 & 60 & 60\\
V384 Per & 535& AMC & 3100& 1200& 0.12 & 8393& 582& 626& 700 & 30 & 30\\
V688 Mon & 653& AMC & 2900& 1000& 0.34 & 40914& 5699& 6622& 1740 & 80 & 80\\
V821 Her & 524& AMC & 2500& 1300& 0.2 & 8165& 556& 597& 700 & 61 & 50\\
V1259 Ori & 696& AMC & 2500& 1300& 0.75 & 11890& 1013& 1108& 1550 & 90 & 90\\
V1426 Cyg & 470& AMC & 2500& 1300& 0.07 & 7070& 435& 463& 600 & 40 & 20\\
V1968 Cyg & 395& AMC & 2600& 1300& 0.55 & 5625& 286& 301& 990 & 80 & 100\\
\\
\textit{S-type}\\
GI Lup & 469& AMC & 2600& 900& 0.01 & 7050& 432& 461& 1190 & 70 & 80\\
IRC$-$10401 & 480& AMS & 2600& 600& 5& 7270& 456& 487& 3470 & 120 & 110\\
NSV 24833 & 418& AMS & 2400& 900& 4& 6053& 328& 347& 3500 & 120 & 140\\
R And & 409& AMS & 3400& 600& 0.06 & 5885& 312& 329& 390 & 30 & 20\\
R Cyg & 427& AMS & 2600& 600& 0.03 & 6226& 346& 366 & 560 & 40 & 30\\
R Gem & 370& AMC & 3200& 1300& 0.03 & 5151& 241& 252& 1000 & 30 & 30\\
R Lyn & 366& AMC & 3200& 1300& 0.03 & 5076& 233& 244 & 1300 & 40 & 50\\
RT Sco & 449& AMS & 2600& 1200& 0.3 & 6655& 391& 415& 710 & 60 & 60\\
S Cas & 608& AMS & 3000& 1000& 0.7 & 9946& 767& 831& 880 & 60 & 50\\
\bottomrule
\multicolumn{7}{l}{\footnotesize $P$: period in days.} \\
\multicolumn{4}{l}{\footnotesize $^*$\,dust type taken from \citet{Ramstedt2006}} \\
\multicolumn{7}{l}{\footnotesize AMC: amorphous carbon.} \\
\multicolumn{7}{l}{\footnotesize AMS: amorphous silicate.} \\
   \multicolumn{9}{l}{\footnotesize $T_\star$ is the stellar temperature; $T_\mathrm{dust}$ the dust temperature; $\tau_{10}$ the optical depth at 10 $\mu$m.} \\
    \multicolumn{9}{l}{\footnotesize $L_\star ^\mathrm{median}$ the median luminosity; $\sigma_{L_\star} ^\mathrm{-/+}$ the lower/upper uncertainties on the luminosity.} \\
    \multicolumn{9}{l}{\footnotesize $r_\mathrm{PL(M)}$ the median distance calculated with our new PL relation;} \\
    \multicolumn{9}{l}{\footnotesize $\sigma_{r,\mathrm{PL(M)}} ^\mathrm{-/+}$ the lower/upper uncertainties on the PL distance.} \\
    \end{tabular}
    \label{table: RT Appendix1}
  \end{center}
\end{table*}
}
{
\renewcommand{\arraystretch}{1.2}
\begin{table*}
  \renewcommand\thetable{B.1}
   
\caption{continued.}
  \begin{center}
      \begin{tabular}{lccccccccccc}
    \toprule
    Source & $P$ & dust$^*$ & $T_\star$ &$T_\mathrm{dust}$& $\tau_{10}$ & $L_\star ^\mathrm{median}\,$& $\sigma_{L_\star} ^\mathrm{-}\,$& $\sigma_{L_\star} ^\mathrm{+}\,$ &$r_\mathrm{PL(M)}$ & $\sigma _{r,\mathrm{PL(M)}}^-$ &$\sigma _{r,\mathrm{PL(M)}}^+$ \\
     & [days] & & [K]& [K] & & [L$_\odot$] & [L$_\odot$] & [L$_\odot$] &[pc]&[pc]&[pc] \\
     \midrule
S Lyr & 438& AMS & 3100& 600& 0.44 & 6447& 369& 391& 2060 & 80 & 80\\
ST Sgr & 400& AMS & 2600& 600& 0.03 & 5711& 294& 310& 720 & 40 & 70 \\
T Cam & 369& AMC & 2500& 1300& 0.04 & 5138& 239& 251& 980 & 30 & 40\\
T Sgr & 396& AMC & 2700& 1000& 0.02 & 5635& 287& 302& 1010 & 50 & 50\\
TT Cen & 462& AMS & 3000& 1200& 0.11 & 6910& 417& 444 & 1510 & 70 & 70\\
VX Aql & 643& AMC & 2700& 700& 0.01 & 10706& 862& 937& 2450 & 110 & 120\\
W And & 397& AMS & 2600& 600& 0.02 & 5660& 289& 305& 350 & 20 & 20\\
W Aql & 479& AMS & 2700& 1200& 0.7 & 7249& 454& 484 & 340 & 31 & 30\\
WY Cas & 482& AMS & 2700& 600& 0.12 & 7300& 460& 490& 1080 & 50 & 50 \\
$\chi$ Cyg & 408& AMS & 2600& 1200& 0.17 & 5862& 309& 326 & 180 & 10 &5\\
\\
\textit{M-type}\\
BW Cam & 628& AMS & 3000& 800& 0.8 & 10376& 820& 891& 1250 & 50 & 60\\
GX Mon & 527& AMS & 3500& 800& 5& 8227& 563& 605& 1430 & 70 & 70\\
GY Aql & 464& AMS & 2900& 1200& 0.3 & 6950& 421& 449 & 410 & 40 & 40\\
IK Tau & 470& AMS & 3000& 900& 0.8 & 7070& 435& 463& 250 & 10 & 10\\
IRC$-$10529 & 675& AMS & 3600& 1200& 4& 11417& 952& 1039& 930 & 70 & 60\\
IRC$-$30398 & 575& AMS & 3100& 1200& 1.4 & 9233& 681& 735& 670 & 50 & 40\\
IRC+10365 & 455& AMS & 3000& 600& 0.7 & 6773& 403& 428& 780 & 30 & 30\\
IRC+40004 & 720& AMS & 2400& 1200& 2.5 & 12435& 1085& 1189& 1160 & 60 & 70\\
KU And & 720& AMS & 2400& 1200& 2.5 & 12435& 1085& 1189& 1170 & 70 & 60\\
NV Aur & 635& AMS & 3500& 1100& 4& 10530& 840& 912& 1325 & 65 & 65\\
R Cas & 430& AMS & 2900& 1200& 0.28 & 6293& 353& 374 &200 &10 & 2\\
R Hor & 408& AMS & 3100& 600& 0.04 & 5862& 309& 326 & 250 & 10 & 10\\
R Leo & 312& AMS & 2900& 1200& 0.11 & 4113& 148& 154& 100 & 5 & 5\\
R LMi & 372& AMS & 2900& 1200& 0.19 & 5191& 244& 256& 320 & 10 & 10\\
T Ari & 314& AMS & 3100& 700& 0.01 & 4145& 151& 157& 290 & 5 & 10\\
T Cep & 388& AMS & 2900& 800& 0.02 & 5487& 273& 287& 150 & 2 & 2\\
TX Cam & 559& AMS & 3000& 1200& 0.7 & 8894& 640& 690 & 410 & 60 & 60\\
U Men & 409& AMS & 2900& 600& 0.04 & 5881& 311& 328 & 330 & 20 & 10\\
WX Psc & 660& AMS & 3400& 1100& 4& 11082& 909& 991& 720 & 30 & 30\\
    \bottomrule
    \end{tabular}
    \label{table: RT Appendix2}
  \end{center}
\end{table*}
}
\begin{figure*}[hbt!]
  \includegraphics[page=1,width=1.00\textwidth]{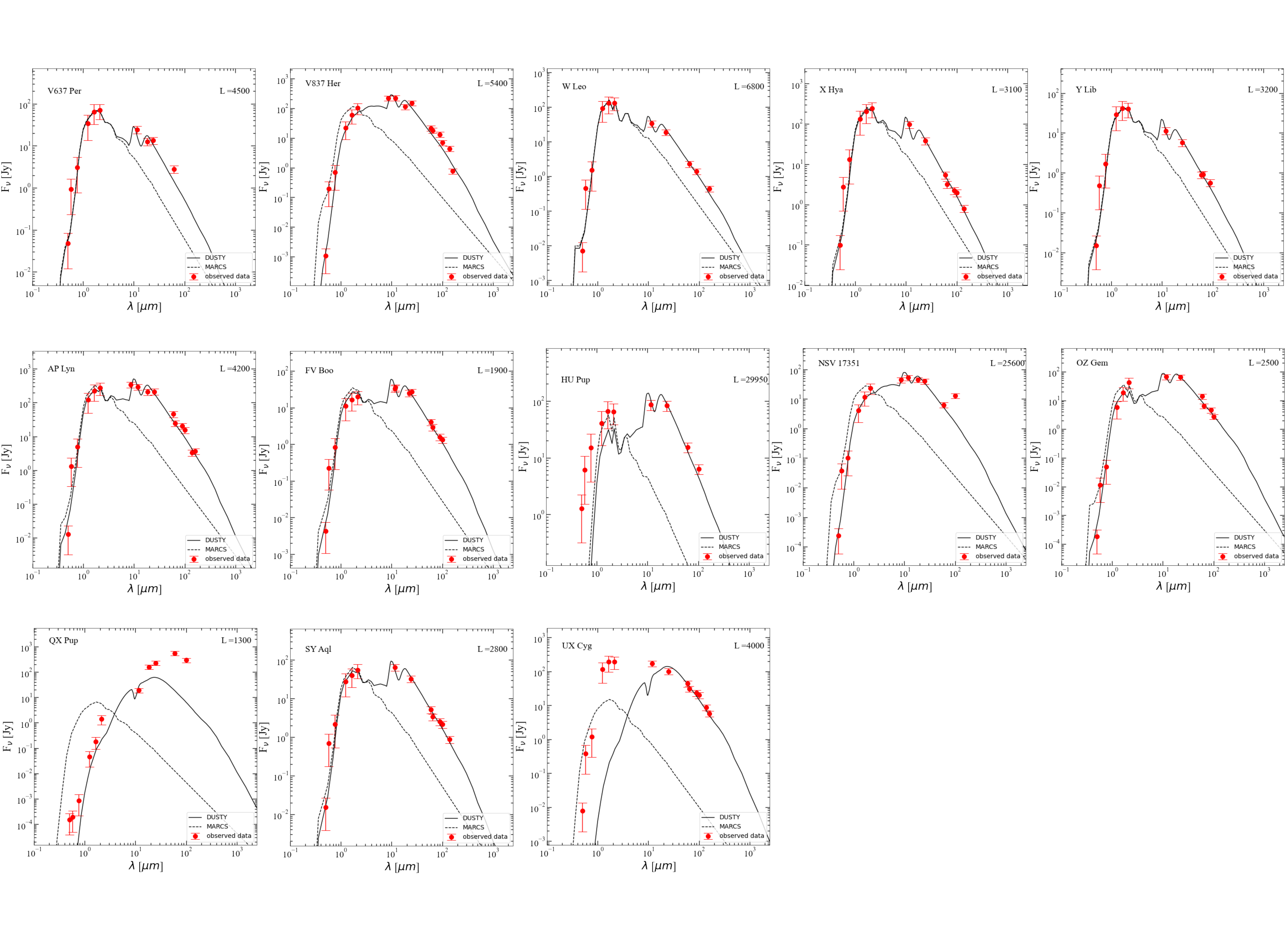}
   
    \caption{SED fitting of the stellar and dust emission with DUSTY for the sources in the VLBI sample. The name of each source is given in the upper left corner of each panel, and the derived luminosity is given in the upper right corner, in solar luminosity. The last two rows correspond to the outliers that were excluded from the PL relation determination (see text).}
    \label{fig:SED2}
\end{figure*}
\begin{figure*}[hb!]
   \includegraphics[page=1,width=1.02\textwidth]{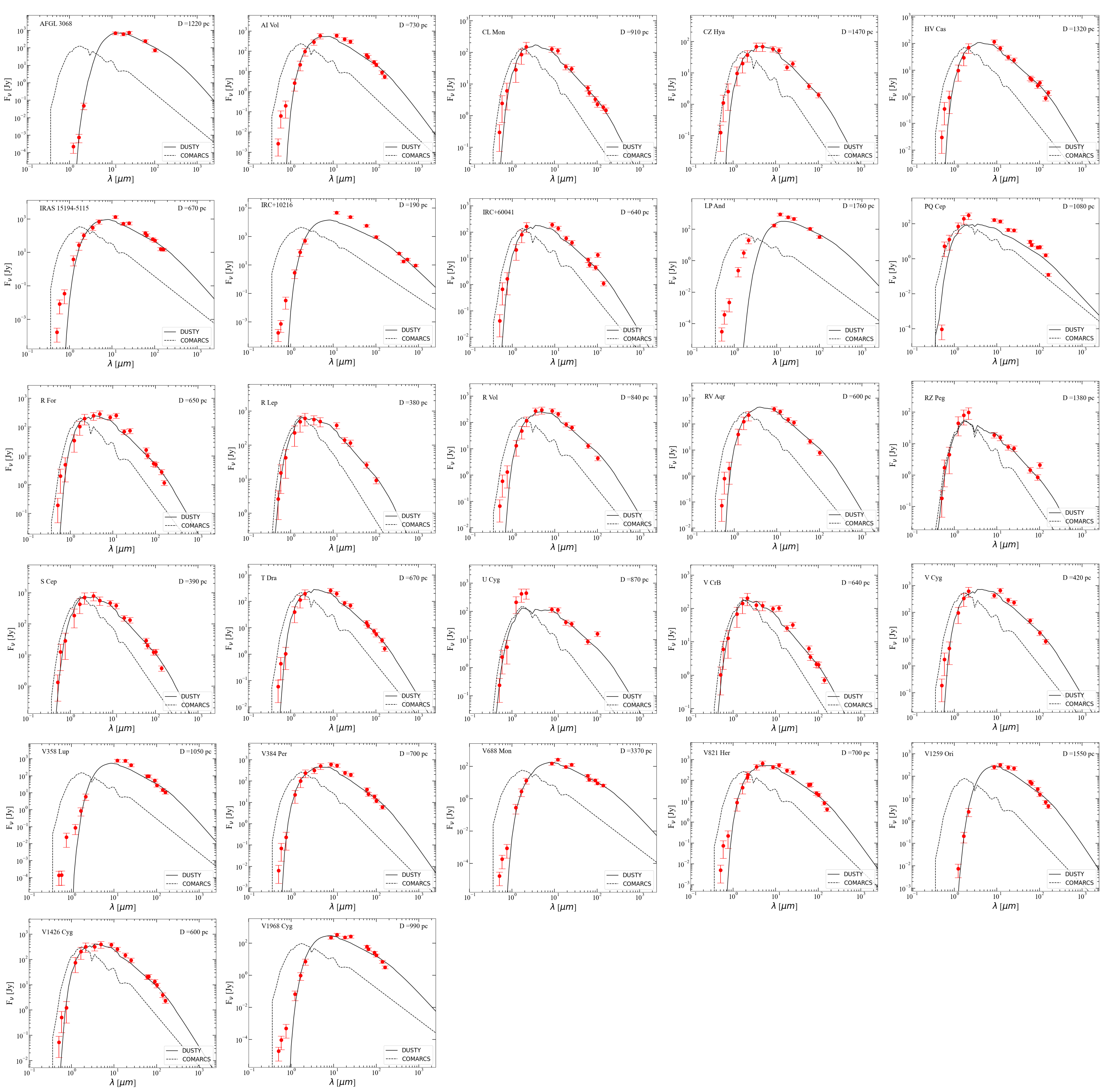}
    
    \caption{SED fitting of the stellar and dust emission of the C-type Miras in the DEATHSTAR sample with DUSTY. The name of each source is given in the upper left corner of each panel, and the derived distance is given in the upper right corner, in pc.}
    \label{fig:SEDextra1}
\end{figure*}
\begin{figure*}[hb!]
   \includegraphics[page=1,width=1.02\textwidth]{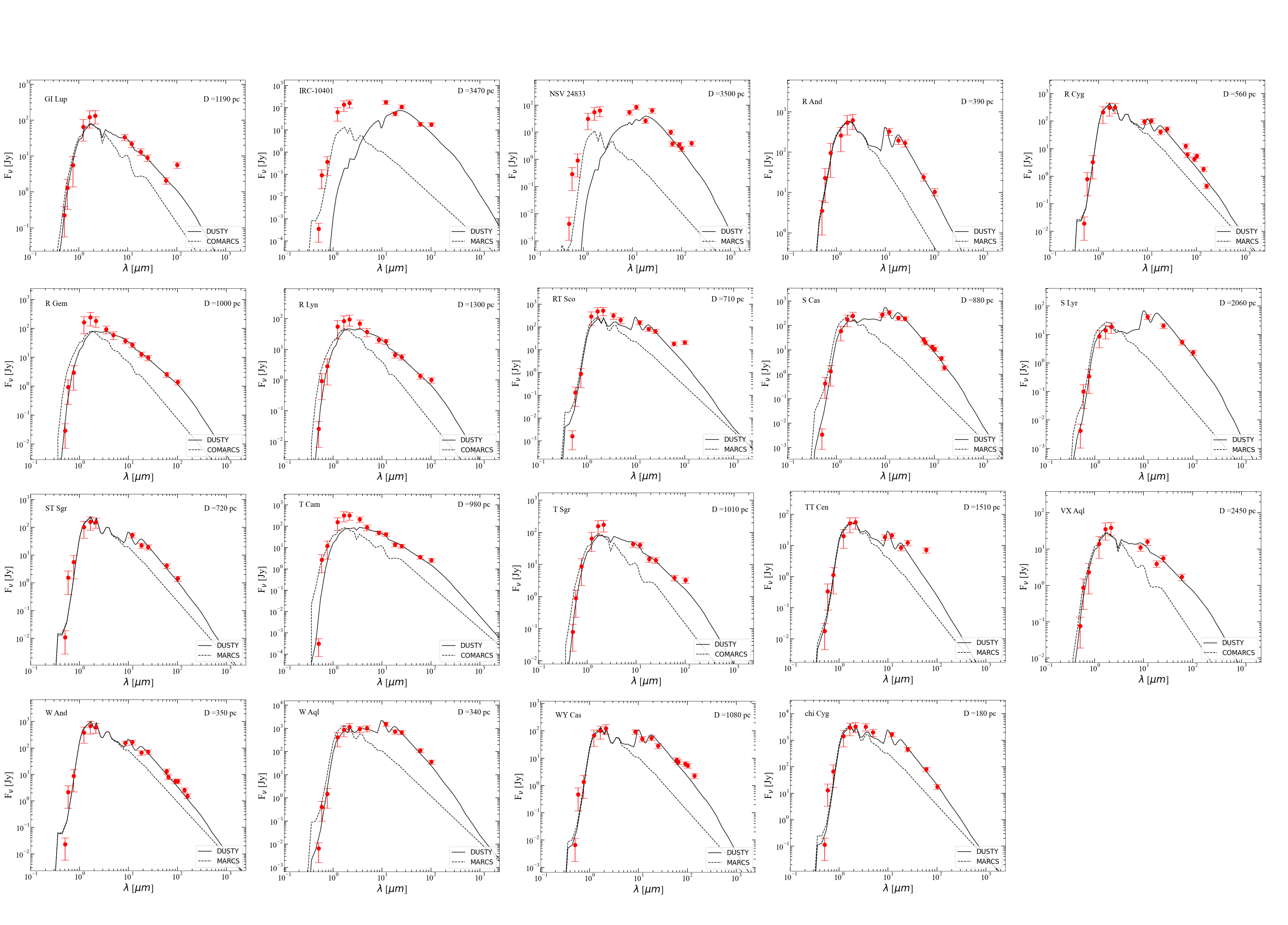}
    
    \caption{SED fitting of the stellar and dust emission of the S-type Miras in the DEATHSTAR sample with DUSTY. The name of each source is given in the upper left corner of each panel, and the derived distance is given in the upper right corner, in pc.}
    \label{fig:SEDextra2}
\end{figure*}
\begin{figure*}[hb!]
   \includegraphics[page=1,width=1.02\textwidth]{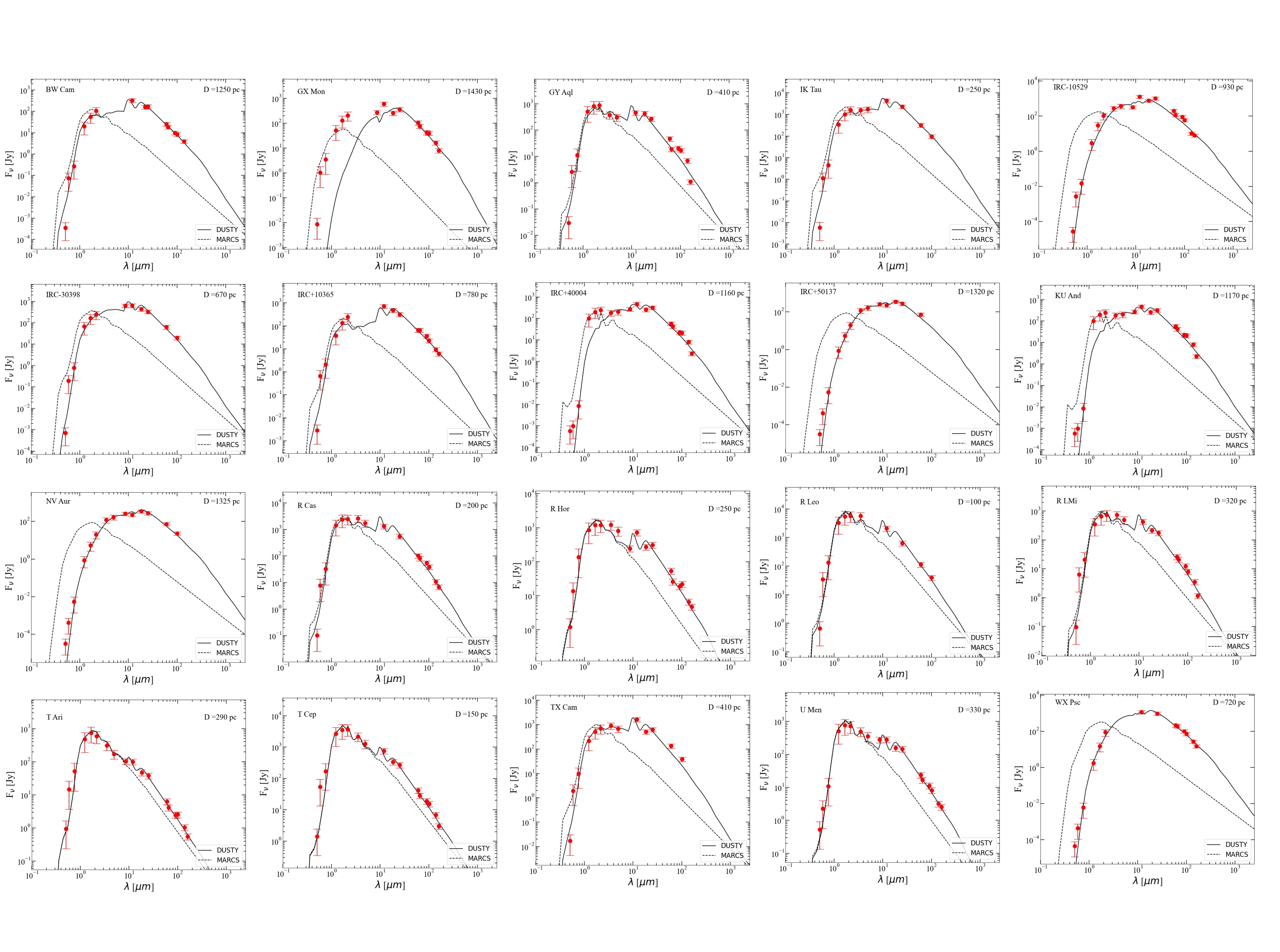}
    
    \caption{SED fitting of the stellar and dust emission for the M-type Miras in the DEATHSTAR sample with DUSTY. The name of each source is given in the upper left corner of each panel, and the derived distance is given in the upper right corner, in pc.}
    \label{fig:SEDextra3}
\end{figure*}
\FloatBarrier
\section{New distance catalogue}
\begin{strip}
{
\begin{minipage}{0.5\textwidth}
\renewcommand{\arraystretch}{1.22}
\captionof{table}{Best distance estimates for the DEATHSTAR sample. }
  \begin{center}
    \begin{tabular}{lccccl}
     \toprule
     Source & $r_\mathrm{median}\,$& $\sigma_\mathrm{r} ^{-} \,$&$\sigma_\mathrm{r} ^{+} \,$& Type & Spectral\\
     & [pc] & [pc] & [pc] & &type\\
     \midrule
  AA Cam & 489 & 60 & 79 & G$_\mathrm{AGB}$ &S \\
  AA Cyg & 440 & 53 & 45 & PL(SRb)&S \\
  AFGL 3068 & 1220 & 60 & 70 & PL(M$_\mathrm{out}$) &C\\
  AH Dra & 391 & 34 & 44 & PL(SRb) &M \\
  AI Vol & 583 & 56 & 70 & G$_\mathrm{AGB}$ & C\\
  AM Cen & 897 & 103 & 134 & G$_\mathrm{AGB}$ &S \\
  AP Lyn & 501 & 10 & 10 & V & M$^*$\\
  AQ And & 768 & 72 & 89 & G$_\mathrm{AGB}$&C \\
  AQ Sgr & 470 & 51 & 57  & PL(SRb)&C \\
  BD+06 319 & 312 & 52 & 80 & G$_\mathrm{AGB}$ & M\\
  BK Vir & 242 & 39 & 59 & G$_\mathrm{AGB}$ & M\\
  BL Ori & 437 & 49 & 52 & PL(SRb) & C\\
  BM Gem & 1243 & 201 & 293 & G$_\mathrm{AGB}$ &C \\
  BW Cam & 1250 & 50 & 60 & PL(M$_\mathrm{out}$)& OH/IR\\
  BX Cam & 579 & 10 & 10 & V &M\\
  BX Eri & 476 & 23 & 25 & V &M\\
  CL Mon & 910 & 30 & 40 & PL(M)&C \\
  CS Dra & 457 & 69 & 99 & G$_\mathrm{AGB}$ &M\\
  CSS2 41 & 3792 & 700 & 497 & G$_\mathrm{AGB}$&S \\
  CW Cnc & 262 & 27 & 35 & G$_\mathrm{AGB}$ & M\\
  CZ Hya & 1514 & 253 & 364 & G$_\mathrm{AGB}$ & C\\
  DK Vul & 843 & 80 & 88 & PL(SRa)&S \\
  DR Ser & 1200 & 223 & 369 & G$_\mathrm{AGB}$ &C\\
  DY Gem & 644 & 76 & 62 & PL(SRa)&S \\
  EP Aqr & 103 & 10 & 13 & PL(SRb)&M \\
  EP Vul & 461 & 45 & 56 & PL(SRb)&S\\
  FU Mon & 795 & 99 & 132 & G$_\mathrm{AGB}$&S \\
  FV Boo & 1034 & 60 & 67 & V& M$^*$\\
  GI Lup & 1019 & 169 & 254 & G$_\mathrm{AGB}$ & S \\
  GX Mon & 1430 & 70 & 70 & PL(M$_\mathrm{out}$) &OH/IR \\
  GY Aql & 410 & 40 & 40 & PL(M) & M\\
  HS UMa & 356 & 11 & 13 & V & M\\
  HU Pup & 3437 & 426 & 511 & V &M \\
  HV Cas & 1205 & 163 & 225 & G$_\mathrm{AGB}$ & C\\
    \bottomrule
 \multicolumn{5}{l}{\footnotesize $r_\mathrm{median} $ is the median distance. } \\
\multicolumn{5}{l}{\footnotesize $\sigma_\mathrm{r} ^{-/+}$the lower/upper distance uncertainty.} \\
 \multicolumn{5}{l}{\footnotesize V: VLBI; G: Gaia. } \\
  \multicolumn{5}{l}{\footnotesize PL(M): PL relation for Miras (this work). } \\
  \multicolumn{5}{l}{\footnotesize PL(M$_\mathrm{out}$): PL(M) outside the period range.} \\
 \multicolumn{5}{l}{\footnotesize PL(SRa/b): PL relation for SRa/b by \citet{Knapp2003}. } \\
  \multicolumn{5}{l}{\footnotesize  (:) indicates that the SR type is unknown. } \\
 \multicolumn{5}{l}{\footnotesize Spectral types: M (C<O), S (C$\sim$O), }\\
 \multicolumn{5}{l}{\footnotesize  C (C>O), OH/IR (see text).}\\
 \multicolumn{5}{l}{\footnotesize M$^*$: suspected OH/IR, U: not defined in SIMBAD. }\\
    \end{tabular}
    \label{table: dist Gaia1}
  \end{center}
\end{minipage}
\begin{minipage}{0.48\textwidth}
\renewcommand{\arraystretch}{1.145}
\renewcommand\thetable{C.1}
\captionof{table}{continued.}
  \begin{center}
    \begin{tabular}{lccccl}
     \toprule
     Source & $r_\mathrm{median}\,$& $\sigma_\mathrm{r} ^{-} \,$&$\sigma_\mathrm{r} ^{+} \,$& Type & Spectral\\
     & [pc] & [pc] & [pc] & &type\\
     \midrule
  IK Tau & 289 & 54 & 88 & G$_\mathrm{AGB}$ &M \\
  IRAS 15194-5115 & 696 & 93 & 129 & G$_\mathrm{AGB}$ &C \\
  IRC+10216 & 190 & 20 & 20 & PL(M$_\mathrm{out}$) &C \\
  IRC+10365 & 519 & 82 & 122 & G$_\mathrm{AGB}$&M \\
  IRC+40004 & 1160 & 60 & 70 & PL(M$_\mathrm{out}$) & M \\
  IRC+60041 & 1300 & 175 & 241 & G$_\mathrm{AGB}$ &C \\
  IRC$-$10401 & 3470 & 120 & 110 & PL(M) &S \\
  IRC$-$10529 & 930 & 70 & 60 & PL(M$_\mathrm{out}$) & OH/IR \\
  IRC$-$30398 & 670 & 50 & 40 & PL(M$_\mathrm{out}$)& M \\
  KU And & 1170 & 70 & 60 & PL(M$_\mathrm{out}$)& M\\
  L$^2$ Pup & 102 & 10 & 14 & PL(SRb) &M \\
  LP And & 428 & 40 & 50 & G$_\mathrm{AGB}$ & C\\
  NP Pup & 586 & 70 & 91 & G$_\mathrm{AGB}$ &C \\
  NSV 17351 & 4064 & 157 & 168 & V & OH/IR\\
  NSV 24833 & 3500 & 120 & 140 & PL(M) & S \\
  NV Aur & 1325 & 65 & 65 & PL(M$_\mathrm{out}$) &M \\
  OH 56.1 +2.1 & 3953 & 616 & 392 & G$_\mathrm{AGB}$ &U \\
  OZ Gem & 1246 & 58 & 63 & V & M$^*$\\
  PQ Cep & 631 & 45 & 54 & G$_\mathrm{AGB}$ & C\\
  QX Pup & 1652 & 78 & 86 & V & OH/IR \\
  R And & 390 & 30 & 20 & PL(M)&S \\
  R Aqr & 220 & 11 & 12 & V & M\\
  R Cas & 200 & 10 & 2 & PL(M) &M \\
  R Cnc & 266 & 19 & 22 & V & M\\
  R Crt & 237 & 40 & 64 & G$_\mathrm{AGB}$ & M\\
  R Cyg & 555 & 45 & 54 & G$_\mathrm{AGB}$ & S\\
  R Dor & 44 & 4 & 5 & PL(SRb)& M \\
  R For & 507 & 24 & 27 & G$_\mathrm{AGB}$ & C \\
  R Gem & 847 & 159 & 256 & G$_\mathrm{AGB}$ & S \\
  R Hor & 260 & 48 & 77 & G$_\mathrm{AGB}$ & M \\
  R Hya & 126 & 2 & 3 & V & M\\
  R LMi & 320 & 10 & 10 & PL(M) &M \\
  R Leo & 100 & 5 & 5 & PL(M)& M\\
  R Lep & 471 & 64 & 88 & G$_\mathrm{AGB}$&C \\
  R Lyn & 880 & 105 & 138 & G$_\mathrm{AGB}$ & S \\
  R Peg & 374 & 36 & 44 & V &M \\
  R Scl & 408 & 61 & 86 & G$_\mathrm{AGB}$&C \\
  R UMa & 508 & 12 & 14 & V &M\\
  R Vol & 662 & 68 & 86 & G$_\mathrm{AGB}$ & C\\
  RR Aql & 411 & 11 & 12 & V &M \\
  RS And & 327 & 34 & 36 & PL(SRa) & M\\
  RS CrA & 1467 & 97 & 112 & G$_\mathrm{AGB}$ & U\\
  RT Cap & 564 & 69 & 92 & G$_\mathrm{AGB}$ & C\\
  RT Sco & 710 & 60 & 60 & PL(M) & S\\
  RT Vir & 227 & 6 & 7 & V & M\\
  RV Aqr & 586 & 50 & 61 & G$_\mathrm{AGB}$ & C \\
  RV Cam & 382 & 41 & 36 & PL(SRb)& M \\
  \bottomrule
    \end{tabular}
  \end{center}
\end{minipage}
}
\end{strip}

{
\renewcommand{\arraystretch}{1.2}
\begin{table}[t]
\renewcommand\thetable{C.1}
 
\caption{continued.}
  \begin{center}
    \begin{tabular}{lrrrrc}
    \toprule
     Source & $r_\mathrm{median}\,$& $\sigma_\mathrm{r} ^{-} \,$&$\sigma_\mathrm{r} ^{+} \,$& Type & Spectral \\
     & [pc] & [pc] & [pc] & & type\\
     \midrule
  RV Cyg & 488 & 49 & 58 & PL(SRb) & C \\
  RW LMi & 319 & 22 & 27 & G$_\mathrm{AGB}$ & C \\
  RW Lep & 636 & 59 & 72 & V & M \\
  RX Boo & 139 & 9 & 11 & V & M \\
  RX Lac & 378 & 51 & 38 & PL(SRb) & S \\
  RY Dra & 401 & 38 & 47 & G$_\mathrm{AGB}$ & C \\
  RY Mon & 875 & 96 & 124 & G$_\mathrm{AGB}$ & C \\
  RZ Peg & 1275 & 128 & 161 & G$_\mathrm{AGB}$ & C \\
  RZ Sgr & 432 & 50 & 66 & G$_\mathrm{AGB}$ & S \\
  S Aur & 1057 & 112 & 133 & PL(SR:) & C \\
  S Cas & 880 & 60 & 50 & PL(M$_\mathrm{out}$)& S \\
  S Cep & 534 & 93 & 144 & G$_\mathrm{AGB}$ & C \\
  S CrB & 424 & 28 & 33 & V & M\\
  S Crt & 433 & 23 & 25 & V & M \\
  S Dra & 253 & 27 & 24 & PL(SRb) & M \\
  S Lyr & 2060 & 80 & 80 & PL(M) & S \\
  S Pav & 184 & 16 & 17 & PL(SRa) & M\\
  S Sct & 438 & 50 & 65 & G$_\mathrm{AGB}$ & C \\
  S Ser & 801 & 25 & 27 & V & M\\
  SS Vir & 583 & 112 & 173 & G$_\mathrm{AGB}$ & C\\
  ST Cam & 625 & 91 & 129 & G$_\mathrm{AGB}$& C \\
  ST Her & 324 & 57 & 86 & G$_\mathrm{AGB}$& S \\
  ST Sco & 412 & 48 & 38 & PL(SRa)& S \\
  ST Sgr & 720 & 40 & 70 & PL(M)& S \\
  SU Vel & 417 & 68 & 103 & G$_\mathrm{AGB}$& M \\
  SV Aqr & 445 & 65 & 90 & G$_\mathrm{AGB}$ & M\\
  SV Peg & 334 & 7 & 7 & V & M\\
  SW Vir & 125 & 16 &12 & PL(SRb) &M \\
  SY Aql & 922 & 56 & 64 & V & OH/IR \\
  SY Scl & 1330 & 50 & 55 & V & M \\
  SZ Car & 689 & 45 & 52 & G$_\mathrm{AGB}$& C \\
  SZ Dra & 470 & 57 & 75 & G$_\mathrm{AGB}$ & M \\
  T Ari & 290 & 5 & 10 & PL(M) & M\\
  T Cam & 980 & 30 & 40 & PL(M) & S \\
  T Cep & 150 & 2 & 2 & PL(M) & M \\
  T Cet & 213 & 24 & 23& PL(SRb) & M \\
  T Dra & 901 & 146 & 214 & G$_\mathrm{AGB}$& C \\
  T Ind & 467 & 38 & 66 & PL(SRb) & C \\
  T Lep & 327 & 4 & 4 & V & M\\
  T Lyr & 427 & 33 & 38 & G$_\mathrm{AGB}$ & C \\
  T Mic & 175 & 15 & 19 & PL(SRb)& M \\
  T Sgr & 1010 & 50 & 50 & PL(M) & S\\
  TT Cen & 1182 & 152 & 208 & G$_\mathrm{AGB}$ & S\\
  TT Cyg & 671 & 43 & 49 & G$_\mathrm{AGB}$& C \\
  TT Tau & 671 & 59 & 73 & G$_\mathrm{AGB}$ & C\\
  TU Gem & 497 & 51 & 57 & PL(SRb)& C \\
  TV Dra & 541 & 90 & 134 & G$_\mathrm{AGB}$ & M\\
    \bottomrule
    \end{tabular}
    \label{table: dist Gaia3}
  \end{center}
\end{table}
}
{
\renewcommand{\arraystretch}{1.2}
\begin{table}[t]
\renewcommand\thetable{C.1}
 
\caption{continued.}
  \begin{center}
    \begin{tabular}{lccccl}
    \toprule
     Source & $r_\mathrm{median}\,$& $\sigma_\mathrm{r} ^{-} \,$&$\sigma_\mathrm{r} ^{+} \,$& Type & Spectral \\
     & [pc] & [pc] & [pc] & & type \\
     \midrule
  TW Hor & 481 & 75 & 109 & G$_\mathrm{AGB}$ & C \\
  TW Oph & 392 & 38 & 46 & PL(SRb) & C \\
  TW Peg & 278 & 48 & 77 & G$_\mathrm{AGB}$ & M\\
  TX Cam & 292 & 35 & 47 & G$_\mathrm{AGB}$ & M\\
  TY Dra & 699 & 89 & 120 & G$_\mathrm{AGB}$ & M \\
  TZ Aql & 524 & 64 & 84 & G$_\mathrm{AGB}$ & M\\
  U Ant & 294 & 40 & 54 & G$_\mathrm{AGB}$ & C\\
  U Cam & 630 & 77 & 102 & G$_\mathrm{AGB}$ & C \\
  U Cyg & 687 & 37 & 41 & G$_\mathrm{AGB}$ & C\\
  U Her & 271 & 19 & 21 & V & M\\
  U Hya & 286 & 34 & 31 & PL(SRb) & C \\
  U Lyn & 792 & 36 & 39 & V & M\\
  U Men & 317 & 28 & 34 & G$_\mathrm{AGB}$ & M\\
  UU Aur & 306 & 30 & 40 & PL(SRb) & C \\
  UX And & 321 & 28 & 41 & PL(SRb) & M\\
  UX Cyg & 1918 & 198 & 250 & V & M\\
  UX Dra & 373 & 37 & 39 & PL(SRb)& C \\
  UY Cen & 718 & 127 & 200 & G$_\mathrm{AGB}$ & S\\
  UY Cet & 449 & 50 & 51 & PL(SRb)& M \\
  V Aql & 379 & 39 & 38 & PL(SRb)& C \\
  V CrB & 836 & 81 & 99 & G$_\mathrm{AGB}$ & C\\
  V Cyg & 420 & 10 & 20 & PL(M)& C \\
  V Hya & 311 & 35 & 36 & PL(SRa)& C \\
  V tel & 423 & 47 & 42 & PL(SRb)& M \\
  V1259 Ori & 1550 & 90 & 90 & PL(M$_\mathrm{out}$) &C \\
  V1302 Cen & 892 & 91 & 114 & G$_\mathrm{AGB}$ & C\\
  V1426 Cyg & 709 & 113 & 168 & G$_\mathrm{AGB}$ & C\\
  V1942 Sgr & 650 & 84 & 115 & G$_\mathrm{AGB}$ & C \\
  V1943 Sgr & 189 & 22 & 19 & PL(SRb) & M \\
  V1968 Cyg & 990 & 80 & 100 & PL(M) & C \\
  V358 Lup & 1050 & 60 & 60 & PL(M$_\mathrm{out}$) & C \\
  V365 Cas & 464 & 48 & 49 & PL(SRb) & S \\
  V384 Per & 700 & 30 & 30 & PL(M$_\mathrm{out}$) & C\\
  V386 Cep & 1037 & 85 & 137 & PL(SRb) &M \\
  V460 Cyg & 360 & 37 & 44 & PL(SRb) & C \\
  V466 Per & 685 & 84 & 112 & G$_\mathrm{AGB}$ & C \\
  V637 Per & 1065 & 22 & 23 & V & M\\
  V688 Mon & 1740 & 80 & 80 & PL(M$_\mathrm{out}$) & C \\
  V821 Her & 700 & 61 & 50 & PL(M$_\mathrm{out}$) & C \\
  V837 Her & 918 & 9 & 8 & V & M\\
  V996 Cen & 578 & 59 & 74 & G$_\mathrm{AGB}$ & C \\
  VX And & 619 & 49 & 59 & G$_\mathrm{AGB}$ & C\\
  VX Aql & 2450 & 110 & 120 & PL(M$_\mathrm{out}$)& S \\
  VY UMa & 444 & 71 & 104 & G$_\mathrm{AGB}$ & C\\
  W And & 350 & 20 & 20 & PL(M) & S\\
  W Aql & 380 & 49 & 68 & G$_\mathrm{AGB}$ & S \\
  W CMa & 777 & 148 & 246 & G$_\mathrm{AGB}$ & C \\
    \bottomrule
    \end{tabular}
    \label{table: dist Gaia4}
  \end{center}
\end{table}
}
{
\renewcommand{\arraystretch}{1.3}
\begin{table}[t]
\renewcommand\thetable{C.1}
 
\caption{continued.}
  \begin{center}
    \begin{tabular}{lccccc}
    \toprule
     Source & $r_\mathrm{median}\,$& $\sigma_\mathrm{r} ^{-} \,$&$\sigma_\mathrm{r} ^{+} \,$& Type & Spectral  \\
     & [pc] & [pc] & [pc] & & type \\
     \midrule
  W Hya & 87 & 9 & 11 & PL(SRa) & M \\
  W Leo & 971 & 18 & 19 & V & M\\
  W Ori & 260 & 24 & 28 & PL(SRb) & C\\
  W Pic & 594 & 43 & 49 & G$_\mathrm{AGB}$ & C\\
  WX Psc & 720 & 30 & 30 & PL(M$_\mathrm{out}$) & OH/IR \\
  WY Cas & 1141 & 196 & 302 & G$_\mathrm{AGB}$ & S \\
  WZ Cas & 465 & 33 & 38 & G$_\mathrm{AGB}$ & C\\
  X Cnc & 332 & 36 & 41 & PL(SRb)& C \\
  X Her & 137 & 23 & 34 & G$_\mathrm{AGB}$ & M\\
  X Hya & 484 & 11 & 12 & V & M\\
  X TrA & 292 & 39 & 28 & PL(SR:) & C \\
  X Vel & 631 & 65 & 82 & G$_\mathrm{AGB}$ & C\\
  XZ Vel & 1151 & 162 & 228 & G$_\mathrm{AGB}$ &C \\
  Y CVn & 249 & 22 & 25 & PL(SRb) & C \\
  Y Hya & 451 & 36 & 65 & PL(SRb)& C\\
  Y Lib & 1173 & 64 & 73 & V & M\\
  Y Lyn & 200 & 21 & 24 & PL(SR:) & S\\
  Y Pav & 381 & 37 & 56 & PL(SRb)& C \\
  Y Scl & 504 & 50 & 54 & PL(SRb) & M \\
  Y Tau & 409 & 43 & 40 & PL(SRb) & C \\
  Y Tel & 442 & 77 & 118 & G$_\mathrm{AGB}$ & M \\
  Y UMa & 348 & 57 & 83 & G$_\mathrm{AGB}$ & M\\
  ZZ Ant & 580 & 54 & 68 & PL(SR:) & S \\
  Z Psc & 660 & 107 & 157 & G$_\mathrm{AGB}$ & C \\
  $\theta$ Aps & 104 & 11 & 13 & PL(SRb) & M\\
  $\chi$ Cyg & 180 & 10 & 5 & PL(M) & S \\
    \bottomrule
    \end{tabular}
    \label{table: dist Gaia5}
  \end{center}
\end{table}
}
\end{appendix}
\end{document}